\documentclass[onecolumn]{aastex631}
\newcommand{\hi}{H\small I}
\newcommand{\hyper}{\small HYPER}
\newcommand{\clfind}{\small CLUMPFIND}
\newcommand{\gclumps}{\small GAUSSCLUMPS}

\shorttitle{Fragmentation of ATLASGAL clumps}
\shortauthors{Pandian et al.}

\begin{document}

\title{Mass assembly in massive star formation: a fragmentation study of ATLASGAL clumps}

\author[0000-0003-4031-1121]{Jagadheep D. Pandian}
\affiliation{Dept. of Earth \& Space Sciences, \\
	Indian Institute of Space Science and Technology, \\
	Trivandrum 695547, India}

\author[0000-0003-0591-8390]{Rwitika Chatterjee}
\affiliation{Space Astronomy Group, \\
	ISITE Campus, U. R. Rao Satellite Centre, \\
	Indian Space Research Organization, \\
	Bengaluru 560037, India}

\author{Timea Csengeri}
\affiliation{Laboratoire d'astrophysique de Bordeaux, \\
	Univ. Bordeaux, CNRS, B18N, \\ 
	allée Geoffroy Saint-Hilaire, \\
	33615, Pessac, France}

\author[0000-0001-5058-695X]{Jonathan P. Williams}
\affiliation{Institute for Astronomy, \\
	University of Hawaii, \\
	2680 Woodlawn Drive, \\
	Honolulu, HI 96822, USA}

\author[0000-0003-4516-3981]{Friedrich Wyrowski}
\affiliation{Max-Planck-Institut f\"{u}r Radioastronomie, \\
	Auf dem H\"{u}gel 69, \\
	53121 Bonn, Germany}

\author[0000-0001-6459-0669]{Karl M. Menten}
\affiliation{Max-Planck-Institut f\"{u}r Radioastronomie, \\
	Auf dem H\"{u}gel 69, \\
	53121 Bonn, Germany}

\begin{abstract}
The mass assembly in star forming regions arises from the hierarchical structure in molecular clouds in tandem with fragmentation at different scales. In this paper, we present a study of the fragmentation of massive clumps covering a range of evolutionary states, selected from the ATLASGAL survey, using the compact configuration of the Submillimeter Array. The observations reveal a wide diversity in the fragmentation properties with about 60\% of the sources showing limited to no fragmentation at the 2\arcsec\ scale, or a physical scale of 0.015 -- 0.09 pc. We also find several examples where the cores detected with the Submillimeter array are significantly offset from the clump potential suggesting that initial fragmentation does not result in the formation of a large number of Jeans mass fragments. The fraction of the clump mass that is in compact structures is seen to increase with source evolution. We also see a significant correlation between the maximum mass of a fragment and the bolometric luminosity of the parent clump. These suggest that massive star formation proceeds through clump fed core accretion with the initial fragmentation being dependent on the density structure of the clumps and/or magnetic fields.
\end{abstract}

\keywords{Molecular clouds(1072) --- Protostars(1302) --- Star formation(1569) --- Submillimeter astronomy(1647)}

\section{Introduction}
Although massive stars play a dominant role in the physical and chemical evolution of the Universe, several aspects of their formation, especially in the very early stages, are still poorly understood. Recent high resolution observations have shown the presence of gravitational unstable disks around massive stars suggesting that massive stars grow in mass by disk accretion similar to the formation of low-mass stars (e.g. \citealt{motogi19, beuther17}). However, the initial conditions at larger spatial scales are still the subject of considerable debate.

Broadly, there are two models for the formation of massive stars: core accretion and competitive accretion \citep{tan14}. In core accretion, the initial fragmentation of a clump creates gravitationally bound cores with a range of masses, with massive stars forming from massive cores with surface densities larger than $\sim 1$~g~cm$^{-2}$ \citep{mckee03}. There is a direct correlation between the mass of an individual core and the mass of stars that form from the core with multiplicity depending on whether or not there is fragmentation in the gravitationally unstable disk \citep{krumholz09}. Thus, the matter that constitutes a massive star is derived primarily from that of the parent core. It also follows that the the stellar initial mass function is related to the core mass function \citep{alves07}.

In competitive accretion, a massive clump fragments into a large number of Jeans mass fragments followed by competitive accretion of matter from the clump reservoir, with the process naturally leading to the stellar initial mass function \citep{bonnell03, bonnell04}. Thus, most of the matter that constitutes the massive star is derived from the entire clump rather than an individual core/fragment \citep{smith09}. The process of competitive accretion also leads to the massive star being located at the center of the gravitational potential of the clump \citep{bonnell01} and naturally results in the formation of massive stars in clusters.

One of the observational techniques to discriminate between the theories is to trace the distribution of mass from large clump scales ($\gtrsim 0.5$~pc) to that of cores ($\lesssim 0.1$~pc). Historically, observational studies targeting the early stages of massive star formation have faced challenges due to their rarity, rapid evolutionary time-scale, and deeply embedded clustered environment. However, the completion of blind surveys of the Galactic plane from millimeter to near-infrared wavelengths has led to a significant advancement in the ability to identify massive star forming regions at different evolutionary stages.

There have been numerous studies targeting the fragmentation of massive clumps at various scales, with the number of studies being too numerous to list comprehensively. These studies reveal a wide diversity in the fragmentation properties with some clumps undergoing limited fragmentation (e.g. \citealt{zhang09, bontemps10, palau13, zhang15, csengeri17, beuther18, figueira18}), while other sources manifest significant fragmentation (e.g. \citealt{bontemps10, palau13, cyganowski17, beuther18, sanhueza19, svoboda19}). Since this diversity provides evidence both in favor of and against both theories, there is still significant debate regarding the mechanism leading to the formation of massive stars.

In this paper, we report on the results of a study targeting the fragmentation of 16 massive clumps spanning a range of evolutionary stages using the Submillimeter Array (SMA) in its compact configuration. We describe the sample selection process and sample characterization in Section~\ref{section:sample}. The SMA observations and data reduction are described in Section~\ref{section:obs}, while Section~\ref{section:results} outlines the results, methodology for identifying individual fragments and measuring their properties along with notes on selected sources. A discussion of the fragmentation observed along with the implications for the formation process is presented in Section~\ref{section:discussion}. Finally, a summary of the results obtained are listed in Section~\ref{section:conclusion}.

\section{Sample Selection}\label{section:sample}
Our sample of massive clumps were selected from the Apex Telescope Large Area Survey of the Galaxy at 870~$\mu$m (ATLASGAL; \citealt{atlasgal}). We selected a total of 16 sources (see Table~\ref{srclist}) spanning a range of evolutionary states, inferred using their properties at mid-infrared wavelengths (Troost, T., Masters thesis). We ensured that the sources could be conveniently grouped on the sky so that three to four sources could be targeted in a single observing track with the SMA.

In order to characterize the large scale properties of the clumps targeted in this study, we constructed the spectral energy distributions (SEDs) of the sources by combining the 870~$\mu$m ATLASGAL data with 500, 350, 250 and 160~$\mu$m data from the Herschel infrared Galactic plane survey (Hi-GAL; \citealt{higal}). We did not use the 70~$\mu$m data from Hi-GAL since the emission at this wavelength has a significant contribution from warm dust, especially in the more evolved sources. We carried out photometry using the IDL based hybrid photometry and extraction routine (\hyper; \citealt{hyper}). The variable background in the Hi-GAL maps is modelled as a 2D polynomial followed by a 2D Gaussian fit to the source. In case of crowded fields, a multi-Gaussian fit is performed for de-blending. The unique feature of multi-wavelength photometry using \hyper\ is that it uses the same aperture, specified at a reference wavelength, for photometry in all bands. This ensures that the same volume of gas and dust is included for flux determination at all wavelengths. We used a reference wavelength of 250~$\mu$m for defining the aperture since the emission morphology and resolution is comparable to that of the ATLASGAL data (18$''$ at 250~$\mu$m versus 18.2$''$ at 870~$\mu$m) while the signal to noise ratio is significantly better (see Appendix~\ref{appendix:hyper} for more details). The \hyper\ flux densities are generally comparable to the Hi-GAL catalogue fluxes except when the sources are affected by blending (especially at 500~$\mu$m). The 870~$\mu$m flux densities determined using \hyper\ are also consistent with the ATLASGAL compact source catalog \citep{csengeri14} determined using the \gclumps\ algorithm \citep{gaussclumps}.

\begin{deluxetable*}{ccccccccccccc}
\tablecaption{Properties of the ATLASGAL clumps \label{srclist}}
\rotate
\tablewidth{0pt}
%\tabletypesize{\scriptsize}
\tablehead{
\colhead{ATLASGAL} & \colhead{Short} & \colhead{$\alpha$} & \colhead{$ \delta $} & \colhead{$v_\mathrm{LSR}$} & \colhead{$d$} & \colhead{$T_d$} & \colhead{$M$} & \colhead{$L$} & \colhead{$R_\mathrm{eff}$} & \colhead{$\Sigma$} & \colhead{$\alpha_\mathrm{vir}$} & \colhead{$\epsilon$}  \\
\colhead{source name} & {source name} & \colhead{(J2000)} & \colhead{(J2000)} & \colhead{(km s$^{-1}$)} & \colhead{(kpc)} & \colhead{(kpc)} & \colhead{($M_\sun$)} & \colhead{($10^3~L_\sun$)} & \colhead{(pc)} & \colhead{(g cm$^{-2}$)} & \colhead{} & \colhead{} 
}
\colnumbers
\startdata
AGAL036.794$-$00.204 & AG36.79$-$0.20  &  $18^h 59^m 13^s.01$  &  $03\degr20\arcmin31.2\arcsec$  &  78.1  &  8.74  &  $< 13.5$  &  $> 1830$  &  $ < 2.28$  &  0.76  &  0.21  &  $< 1.28$ & $< 0.012$ \\
AGAL036.826$-$00.039 & AG36.83$-$0.04  &  $18^h 58^m 41^s.28$  &  $03\degr26\arcmin49.2\arcsec$  &  60.2  &  3.58  &  10.5  &  460  &  1.23  &  0.24  &  0.53  &  0.98 & 0.025 \\
AGAL036.899$-$00.409 & AG36.90$-$0.41  &  $19^h 00^m 08^s.64$  &  $03\degr20\arcmin33.0\arcsec$  &  80.0  &  8.59  &  16.1  &  1420  &  2.86  &  0.62  &  0.25  &  1.26 & 0.049 \\
AGAL041.049$-$00.247 & AG41.05$-$0.25  &  $19^h 07^m 12^s.72$  &  $07\degr06\arcmin24.5\arcsec$  &  66.0  &  8.42  &  21.1  &  1030  &  13.5  &  0.58  &  0.20  &  1.07 & 0.10  \\
AGAL041.077$-$00.124 & AG41.07$-$0.12  &  $19^h 06^m 48^s.93$  &  $07\degr11\arcmin13.6\arcsec$  &  63.3  &  8.61  &  20.7  &  1280  &  10.4  &  0.82  &  0.13  &  1.32 & 0.075 \\
AGAL046.174$-$00.524 & AG46.17$-$0.53  &  $19^h 17^m 49^s.56$  &  $11\degr31\arcmin06.2\arcsec$  &  50.1  &  3.40  &  14.4  &  156  &  0.33  &  0.14  &  0.53  &  1.13 & 0.028 \\
AGAL046.426$-$00.237 & AG46.43$-$0.24  &  $19^h 17^m 16^s.73$  &  $11\degr52\arcmin30.7\arcsec$  &  52.3  &  3.60  &  12.9  &  370  &  1.02  &  0.31  &  0.26  &  1.18 & 0.041 \\
AGAL047.031+00.244 & AG47.03$+$0.24  &  $19^h 16^m 41^s.66$  &  $12\degr38\arcmin07.1\arcsec$  &  54.9  &  7.45  &  19.3  &  1240  &  12.2  &  0.66  &  0.19  &  1.70 & 0.13 \\
AGAL047.051+00.251 & AG47.05$+$0.25  &  $19^h 16^m 41^s.78$  &  $12\degr39\arcmin22.3\arcsec$  &  56.0  &  7.33  &  14.6  &  1500  &  4.48  &  0.50  &  0.40  &  0.93 & 0.028 \\
AGAL048.606+00.022 & AG48.61$+$0.02  &  $19^h 20^m 31^s.19$  &  $13\degr55\arcmin24.7\arcsec$  &  17.6  &  9.83  &  40.5  &  1330  &  230  &  0.36  &  0.68  &  0.80 & 1.8 \\
AGAL049.253$-$00.411 & AG49.25$-$0.41  &  $19^h 23^m 21^s.19$  &  $14\degr17\arcmin21.5\arcsec$  &  66.1  &  5.41  &  15.7  &  764  &  1.25  &  0.43  &  0.27  &  0.76 & 0.020 \\
AGAL049.599$-$00.249 & AG49.60$-$0.25  &  $19^h 23^m 26^s.59$  &  $14\degr40\arcmin19.9\arcsec$  &  56.7  &  5.41  &  18.3  &  636  &  3.20  &  0.33  &  0.39  &  1.49 & 0.11 \\
AGAL050.284$-$00.391 & AG50.28$-$0.39  &  $19^h 25^m 18^s.09$  &  $15\degr12\arcmin26.8\arcsec$  &  16.1  &  9.53  &  34.2  &  1000  &  190  &  0.35  &  0.54  &  0.71 & 4.3 \\
AGAL052.847$-$00.662 & AG52.85$-$0.66  &  $19^h 31^m 23^s.89$  &  $17\degr19\arcmin49.1\arcsec$  &  48.2  &  5.48  &  22.6  &  490  &  7.40  &  0.25  &  0.52  &  1.20 & 0.023 \\
AGAL052.922+00.414 & AG52.92$+$0.41  &  $19^h 27^m 35^s.33$  &  $17\degr54\arcmin40.3\arcsec$  &  44.7  &  6.22  &  15.6  &  880  &  4.40  &  0.38  &  0.41  &  1.26 & 0.12 \\
AGAL053.141+00.069 & AG53.14$+$0.07  &  $19^h 29^m 17^s.61$  &  $17\degr56\arcmin19.6\arcsec$  &  21.7  &  1.59  &  20.8  &  126  &  1.80  &  0.07  &  1.71  &  0.68 & 0.35 \\
\enddata
\tablecomments{The columns show (1) the source name in the ATLASGAL compact source catalog (2) short name adopted in the paper (3)--(4) J2000 epoch equatorial coordinates (5) radial velocity of the source with respect to the local standard of rest (6) distance to the source (7) dust temperature (8) mass of the clump (9) bolometric luminosity (10) effective radius (11) mass surface density (12) virial parameter and (13) evolutionary state parameter.}
\end{deluxetable*}

\begin{figure}[htb!]
\centering
  \includegraphics[width=0.5\textwidth]{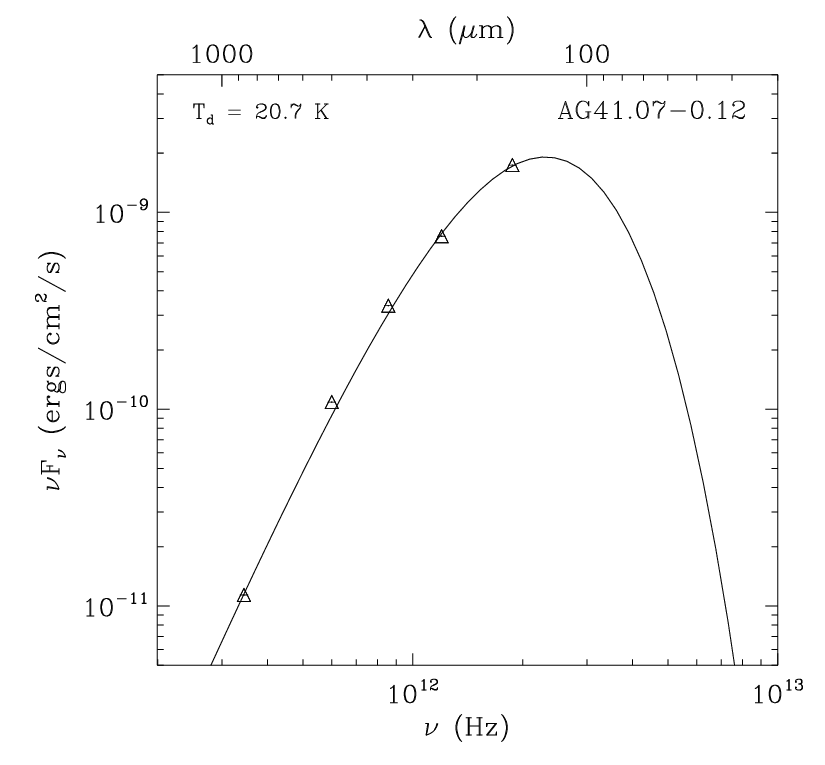}
  \caption{Grey body SED fit to the source AG41.07$-$0.12.}
  \label{samplesedfit}
\end{figure}

\begin{figure*}[htb!]
  \plotone{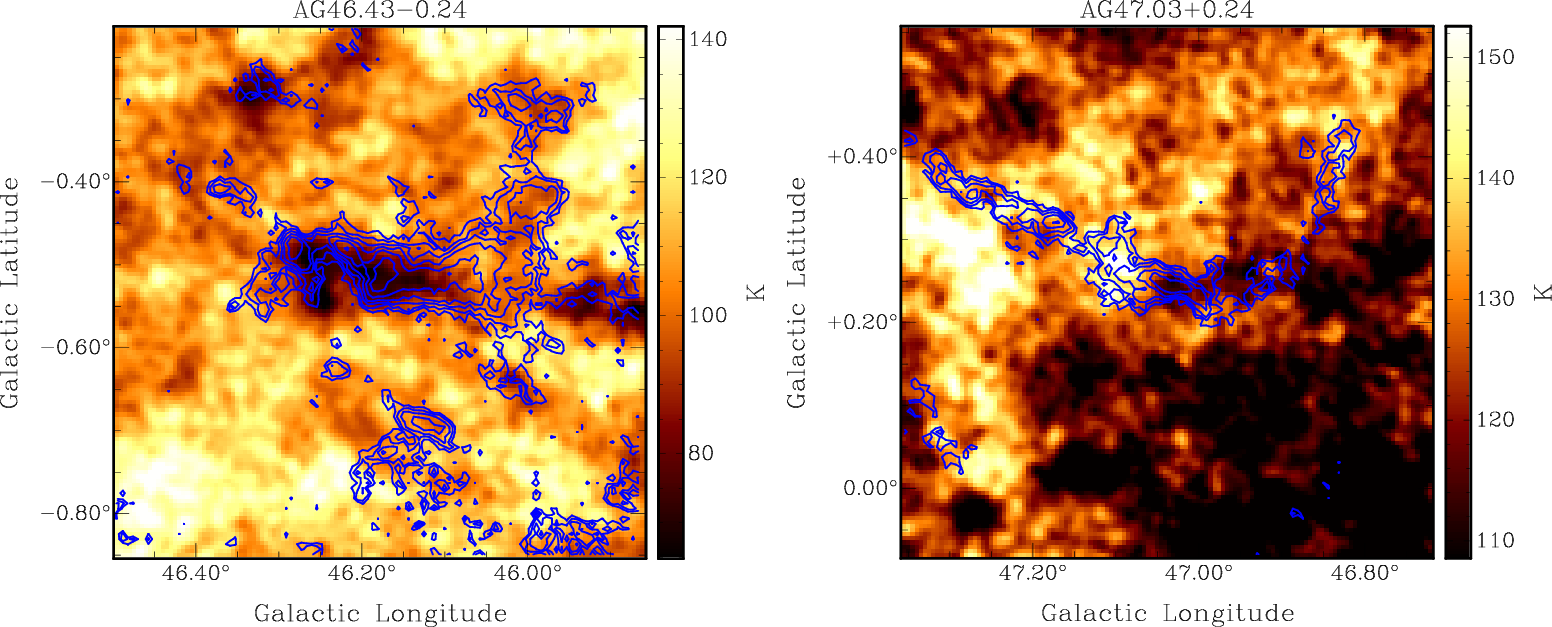}
  \caption{The panels show \hi~ emission from VGPS at the radial velocity of the source with $^{13}$CO contours from GRS overlaid in blue. The left and right panels show examples of sources with and without \hi~ self-absorption respectively.}
    \label{hiselfabs}
\end{figure*}

The dust temperature ($T_d$) was determined by fitting the SED with a single component grey body
\begin{equation}
S_\nu = \Omega B_\nu(T_d)\left[ 1 - \exp\left(-\tau_0 (\nu/\nu_0)^\beta\right) \right]
\end{equation}
where $S_\nu$ is the flux density, $\Omega$ is the solid angle of the source, $B_\nu$ is the black body function, $\tau_0$ is the dust optical depth at reference frequency $\nu_0$ and $\beta$ is the dust emissivity which was allowed to vary between 1 and 3. Fig.~1 shows an example of SED fits for the sources in our sample. The source AG36.79--0.20, which is dark at 70~$\mu$m, is affected by the presence of a second source that is prominent at wavelengths shorter than 250~$\mu$m. Since it was not possible to de-blended the two sources, the SED was fit to the 870 -- 350~$\mu$m data using the constraint that the fit should give a 70~$\mu$m flux density lower than the completeness limit of the Hi-GAL survey.

The isothermal clump mass was determined from the 870~$\mu$m flux using
\begin{equation}
M = \frac{S_\nu d^2 R}{\kappa_\nu B_\nu(T_d)}
\end{equation}
where $R$ is the gas to dust ratio that is assumed to be 100, $d$ is the distance to the source, and $\kappa_\nu$ is the dust opacity which is taken to be 1.85~cm$^2$~g$^{-1}$ \citep{ossen94}. We have used the kinematic distance for all sources except for those in the W51 region (AG49.25--0.41 and AG49.60--0.25) for which we have adopted the distance measured using trigonometric parallax towards H$_2$O masers \citep{sato10}. The kinematic distances were determined using the Galactic rotation curve prescribed by \citet{reid14} adopting the radial velocity of NH$_3$~(1,1) \citep{wienen12}. The kinematic distance ambiguity was resolved by looking for \hi\ absorption (if radio continuum is associated with the source or region) or self-absorption (if radio continuum is absent) using data from the VLA Galactic Plane Survey (VGPS; \citealt{vgps}) and the Boston University -- Five College Radio Astronomy Observatory Galactic ring survey (GRS; \citealt{grs}). The resolution of the kinematic distance ambiguity using \hi\ self-absorption is challenging since the natural line profile of \hi\ can be mistaken for an absorption feature. To avoid incorrect resolution of the distance ambiguity, we compared the morphology of $^{13}$CO emssion in GRS with that of \hi\ and only classified a source at the near distance if the morphology of absorption in \hi\ matched with that of $^{13}$CO emssion. Fig.~\ref{hiselfabs} shows examples of sources determined to be at near and far distances using this technique. In the case of source AG47.03+0.24 (Fig.~\ref{hiselfabs}), while there appears to be a \hi\ self-absorption feature at the source centre, the morphology of the absorption feature does not match with that of the CO emission showing this source to be at the far distance.

The bolometric luminosity of the clumps can be determined from the total flux by integrating the SED. The SED above from sub-millimetre to far-infrared wavelengths is sufficient to characterize sources that are in very early stages of evolution (class A sources) that are dark at 70~$\mu$m. However, sources at later evolutionary stages have significant emission up to mid-infrared or even near-infrared wavelengths. In order to determine the bolometric luminosity of these sources, we extended the SED to shorter wavelengths using data from MIPSGAL, MSX \citep{msx}, GLIMPSE \citep{glimpse} and 2MASS \citep{twomass} surveys. We then fit the SEDs using the young stellar object (YSO) models of \citet{robitaille}. The objective of this fitting is only to determine the bolometric luminosity of the source and is hence not expected to be sensitive to the precise details of the models themselves.

The effective radius of the source is computed from the {\small HYPER} 2D Gaussian fit using
\begin{equation}
R_\mathrm{eff} = \sqrt{\frac{A}{\pi}} = d\left[ \frac{\sqrt{(\theta_\mathrm{maj}^2 - \theta_b^2)(\theta_\mathrm{min}^2 - \theta_b^2)}}{4\log 2} \right]^{1/2}
\end{equation}
where $A$ is the deconvolved area, $\theta_\mathrm{maj}$ and $\theta_\mathrm{min}$ are the full width at half maximum (FWHM) major and minor axes respectively and $\theta_b$ is the FWHM beam width of the telescope (18\arcsec~for the Hi-GAL 250~$\mu$m data). The surface density is determined from the mass and effective radius using
\begin{equation}
\Sigma = \frac{M}{\pi R_\mathrm{eff}^2}
\end{equation}
The gravitational stability of the clumps against collapse can be inferred from the virial parameter which is defined as
\begin{equation}
\alpha_\mathrm{vir} = \frac{5\sigma^2 R_\mathrm{eff}}{GM}
\end{equation}
where $\sigma$ is the velocity dispersion. We have calculated the virial parameter using the velocity dispersion measured from the GRS $^{13}$CO data. The $^{13}$CO spectrum is extracted at the ATLASGAL source position (listed in Table~\ref{srclist}) and fit with a Gaussian to determine the velocity dispersion. Since $^{13}$CO is a trace molecule, the velocity dispersion that characterizes the total column of gas \citep{fuller92} is computed using
\begin{equation}
\sigma^2 = \sigma_\mathrm{CO}^2 + \frac{kT}{m_\mathrm{H}}\left(\frac{1}{\mu} - \frac{1}{\mu_\mathrm{CO}}\right)
\end{equation}
where $T$ is the kinetic temperature of the gas and is assumed to be the dust temperature ($T_d$) and $\mu$ is the mean molecular weight of the medium and is assumed to be 2.35 (derived from the cosmic abundance scale of \citealt{cosmicabundance}).

Table~\ref{srclist} shows that the targeted sample of ATLASGAL clumps are potential sites for forming massive stars. The clump masses are typically significantly higher than 100~$M_\odot$ and the surface densities are well above 0.05~cm$^2$~g$^{-1}$, which is considered as an empirical lower limit for forming massive stars \citep{urquhartmethanol}. The clump masses and effective radius also satisfy the empirical relation of \citet{kauffmann10} highlighting their potential to form massive stars. The virial parameter ranges from 0.68 to 1.71 showing that almost all sources are unstable to gravitational collapse. The dust temperature ranges from 10.5 to 40.5~K showing that the sources span a range of evolutionary states. 

\begin{figure}[htb!]
	\centering
	\includegraphics[width=0.5\textwidth]{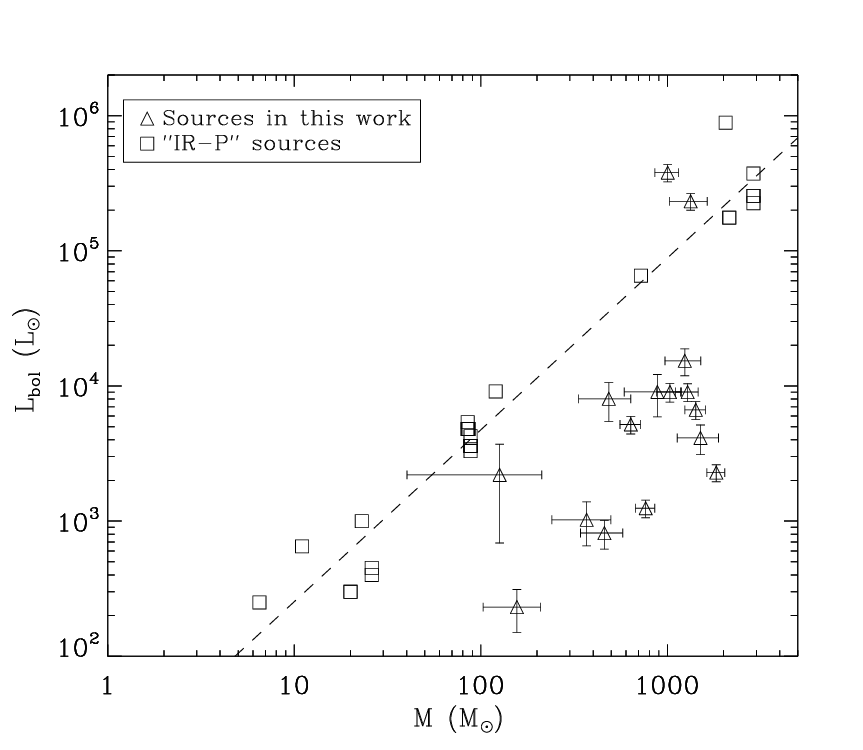}
	\caption{The $L-M$ diagram of the sample. The sources targeted in this work are shown in triangles while the `IR-P' sources of \citet{molinari08} are shown in squares. The dashed line represents the fit to `IR-P' sources with sources below the dashed line having $\epsilon > 1$.}
	\label{lmdiagram}
\end{figure}

One can obtain a more quantitative estimate of evolutionary state of the clumps from their mass to light ratio. Using the turbulent core model of \citet{mckee03}, \citet{molinari08} derived evolutionary tracks for massive young stellar objects (MYSOs) in the $L_\mathrm{bol}-M_\mathrm{env}$ diagram. Although the tracks have been derived from the turbulent core model, the overall features are expected to be model independent. This is because accretion with minimal loss of the envelope mass is expected to be the dominant process prior to the central object reaching the zero age main sequence (ZAMS), while gas dispersion or envelope clearing is expected to dominate the later phases. This results in vertical evolutionary tracks in the $L-M$ diagram during the accretion phase with the luminosity dominated by accretion luminosity, while the late phase is characterized by horizontal tracks towards lower masses since the luminosity is expected to be constant post ZAMS. Based on a study of MYSOs in 42 regions of massive star formation, \citet{molinari08} found that the massive sources with SEDs consistent with the presence of an embedded ZAMS star (called ``IR-P'' sources) were roughly at the transition between the accretion and clearing phase (see Fig.~9 of \citealt{molinari08}). We thus fit the bolometric luminosity as a function of envelope mass for the ``IR-P'' sources of \citet{molinari08} and use this fit to define an evolutionary state parameter
\begin{equation}
\epsilon = 0.074 \left(\frac{L}{L_\sun}\right)\left(\frac{M}{M_\sun}\right)^{-1.27}
\end{equation}\label{eq:mlratio}
The value of unity for $\epsilon$ defines the fit of luminosity vs. mass for the ``IR-P'' sources, with $\epsilon < 1$ in the accretion phase and $\epsilon > 1$ in the clearing phase. One thus expects $\epsilon$ to increase with time as the source evolves. We also note that while one can use $(L/M)$ directly as a marker for evolutionary stage, we prefer to use $\epsilon$ since it is based on the models developed by \citet{molinari08}. Fig.~\ref{lmdiagram} shows the clumps targeted in this study in the $L-M$ diagram with the evolutionary state parameter being tabulated in Table~\ref{srclist}. The value of $\epsilon$ ranges from 0.012 to 4.3. This clearly demonstrates that the clumps span a wide range of evolutionary states with a majority of the sources being in the accretion phase. The source sample is thus well suited to address the question of mass assembly in massive star forming regions.

\section{Observations}\label{section:obs}
The observations towards the 16 ATLASGAL sources were carried out using the SMA in its compact configuration between 2010 May 5 and 2011 October 18. Two sources, AG36.90--0.41 and AG41.05--0.25 were followed up with the SMA in its extended configuration on 2011 August 16. The details of the sources observed and calibrators used are tabulated in Table~\ref{smaobs}.

\begin{deluxetable*}{cccccccc}
\tablecaption{Details of SMA observations \label{smaobs}}
\tablewidth{0pt}
\tabletypesize{\scriptsize}
\tablehead{
\colhead{Source name} & \colhead{Date} & \colhead{Configuration} & \colhead{Antennas} & \colhead{Passband} & \colhead{Gain} & \colhead{Flux} & \colhead{$1\sigma$ noise} \\
\colhead{} & \colhead{} & \colhead{} & \colhead{} & \colhead{calibrator} & \colhead{calibrator} & \colhead{calibrator} & \colhead{(mJy beam$^{-1}$)}
}
\colnumbers
\startdata
AG36.79$-$0.20 & 2010 May 5 & compact & 6 & 3C279 & 1751+096, 1925+211 & Callisto & 1.8 \\
AG36.83$-$0.04 & 2010 May 5 & compact & 6 & 3C279 & 1751+096, 1925+211 & Callisto & 2.4 \\
AG36.90$-$0.41 & 2010 May 5 & compact & 6 & 3C279 & 1751+096, 1925+211 & Callisto & 3.1 \\
AG36.90$-$0.41 & 2011 August 16 & extended & 8 & 3C454.3 & 1751+096, 1925+211 & Ganymede & 2.0\tablenotemark{a} \\
AG41.05$-$0.25 & 2010 June 12 & compact & 7 & 3C279 & 1751+096, 2025+337 & Neptune & 3.1 \\
AG41.05$-$0.25 & 2011 August 16 & extended & 8 & 3C454.3 & 1751+096, 1925+211 & Ganymede & 2.3\tablenotemark{a} \\
AG41.07$-$0.12 & 2010 June 12 & compact & 7 & 3C279 & 1751+096, 2025+337 & Neptune & 2.5 \\
AG46.17$-$0.24 & 2010 June 12 & compact & 7 & 3C279 & 1751+096, 2025+337 & Neptune & 2.6 \\
AG46.43$-$0.24 & 2010 October 4 & compact & 7 & 3C454.3 & 1751+096, 2025+337 & Callisto & 3.3 \\
AG47.03+0.24 & 2010 October 4 & compact & 7 & 3C454.3 & 1751+096, 2025+337 & Callisto & 3.4 \\
AG47.05+0.25 & 2010 October 4 & compact & 7 & 3C454.3 & 1751+096, 2025+337 & Callisto & 4.0 \\
AG48.61+0.02 & 2011 September 23 & compact & 6 & 3C84 & 1751+096, 2025+337 & Callisto & 6.1 \\
AG49.25$-$0.41 & 2010 October 4 & compact & 7 & 3C454.3 & 1751+096, 2025+337 & Callisto & 2.9 \\
AG49.60$-$0.25 & 2011 October 18 & compact & 7 & 3C84 & 1751+096, 2025+337 & Callisto & 6.5 \\
AG50.28$-$0.39 & 2011 September 23 & compact & 6 & 3C84 & 1751+096, 2025+337 & Callisto & 5.7 \\
AG53.14+0.07 & 2011 September 23 & compact & 6 & 3C84 & 1751+096, 2025+337 & Callisto & 4.7 \\
AG52.85$-$0.66 & 2011 October 18 & compact & 7 & 3C84 & 1751+096, 2025+337 & Callisto & 4.5 \\
AG52.92+0.41 & 2011 October 18 & compact & 7 & 3C84 & 1751+096, 2025+337 & Callisto & 6.8 \\
\enddata
\tablenotetext{a}{Noise obtained by imaging compact and extended configurations together}
\end{deluxetable*}

All the observations were carried out using the 345~GHz receiver and the ASIC correlator. The correlator was used in the 1 receiver 4~GHz mode which provides a bandwidth of 4~GHz per sideband. The local oscillator (LO) tuning of the 2010 observations was such that rest frequencies of 333.5 to 337.5~GHz were covered in the lower sideband (LSB) while the upper sideband (USB) covered rest frequencies from 345.5 to 349.5~GHz. The 2011 observations were set up such that rest frequencies of 342 to 346~GHz were covered in LSB, while the USB covered rest frequencies from 354 to 358~GHz. In all observations, the correlator was set up with a uniform channel spacing of 0.8125~MHz ($\sim 0.7$~km~s$^{-1}$) across the entire band.

The data were calibrated using the {\small IDL} based {\small MIR} package. After initial inspection and flagging, the data were weighted by the system temperatures, which ranged from 200 to 600~K depending on elevation. The data were then calibrated for baseline dependent passbands, antenna and time dependent gain and flux using appropriate calibrators. When Neptune was used for flux calibration, the zero spacing flux was used rather than adopting a black body model due to the planet being affected by CO absorption lines. The flux calibration is estimated to be accurate to about 15 per cent. The continuum was then constructed from the line-free channels independently for each sideband. The data were then imported to {\small CASA} (version 4.2) and imaged using the task `clean' using multi-frequency synthesis. The $uv-$range of the SMA in the compact configuration ranges from $\sim 10-70$~m, providing a synthesized beam of around 2\arcsec, with the 1$\sigma$ noise ranging from 1.8 to 6.8~mJy~beam$^{-1}$ (see Table~\ref{smaobs}). The relatively high noise values in some sources is due to the presence of strong emission where the data were limited by dynamic range. The uncertainty in absolute astrometry is estimated to be $\lesssim 0.5\arcsec$ by comparing the positions of the gain calibrators with their catalog positions.

The data from the extended array, which have a $uv-$range of $\sim 25-230$~m, were combined with that of the compact array for both AG36.90--0.41 and AG41.05--0.25 and then imaged. This preserved the total flux density observed in the compact array, but provided the angular resolution of the extended array, which was around $0.6\arcsec \times 0.9\arcsec$. The shortest baseline of the compact array observations was typically lower than 15~k$\lambda$, implying that the observations are sensitive to a largest angular scale of $\sim 13\arcsec$. The beam size of an SMA antenna is $25.5\arcsec$ at the 20\% level and $\sim 30\arcsec$ at the 10\% level, which sets the field of view of the array. 

\section{Results}\label{section:results}
\begin{figure*}[h]
	\centering
	\includegraphics[width=0.78\textwidth]{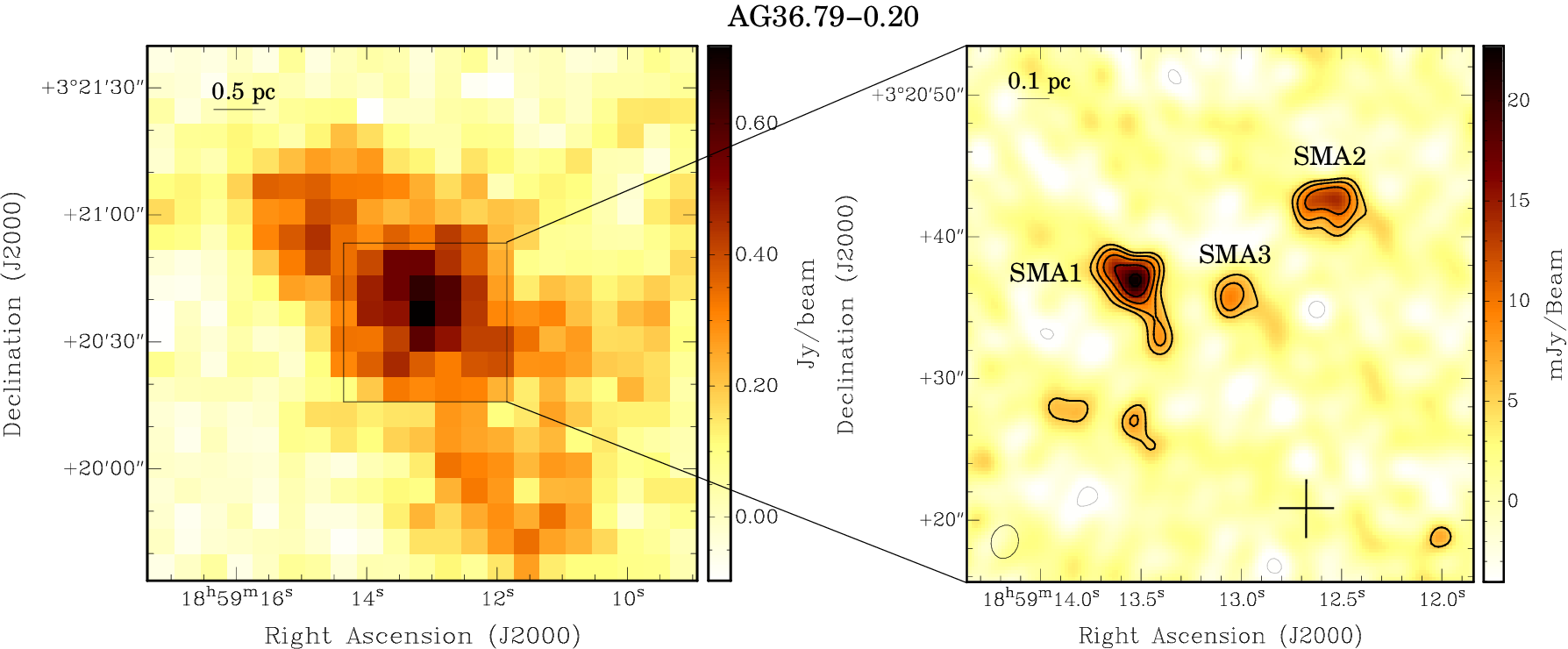}
	\caption{The left panel shows the 870~$\mu$m image from ATLASGAL while the right panel shows the 850~$\mu$m image using the SMA in its compact configuration. The grey contours show emission at the $-3\sigma$ level, while the black contours begin at $3\sigma$ and increase in factors of $\sqrt{2}$. The synthesized beam is shown in the bottom left corner of the SMA image. The field of view of the SMA map in the right panel is shown by the box in the left panel. The names of the fragments detected (see Table~\ref{fraglist}) are indicated in the SMA map. The plus signs indicate the locations of 24~$\mu$m point sources in the MIPSGAL survey. The images for the other sources are available in Appendix~\ref{smafigures}. \label{labocasma}}
\end{figure*}

We detected continuum emission from the SMA observations towards all sources at greater than 5$\sigma$ level. Figure~\ref{labocasma} shows an example of the 870~\micron\ image from ATLASGAL alongside that from the SMA in its compact configuration. The full set of figures are available in Appendix~\ref{smafigures}. Clearly, the maps at high resolution show a range of morphologies, with some sources fragmenting into many cores, while others are dominated by a single core.

The SMA images reveal a diversity in morphology that is a direct consequence of the hierarchical structure in molecular clouds. The hierarchical structure can be analyzed statistically (e.g. through principal component analysis; \citealt{heyer04}) or through segmentation by dividing the data into different structures. Two prominent algorithms used for the latter analysis are \clfind\ \citep{clumpfind} and \gclumps\ \citep{gaussclumps,kramer98}. An alternate method to describe hierarchical structure is through dendrograms \citep{houlahan92, roso08}. The dendrogram technique identifies local maxima as \textit{leaves} which merge into a larger structures at \textit{nodes}. The two main parameters that control the identification of these structures are (a) the height of a leaf above a node (i.e. the difference between the peak of a local maximum and the level at which it merges into a larger structure) and (b) the number of image pixels that are larger than its neighbors over a specified box size. The two parameters are used to control spurious local maxima that result from noise in real data.

\begin{figure*}[h]
	\plotone{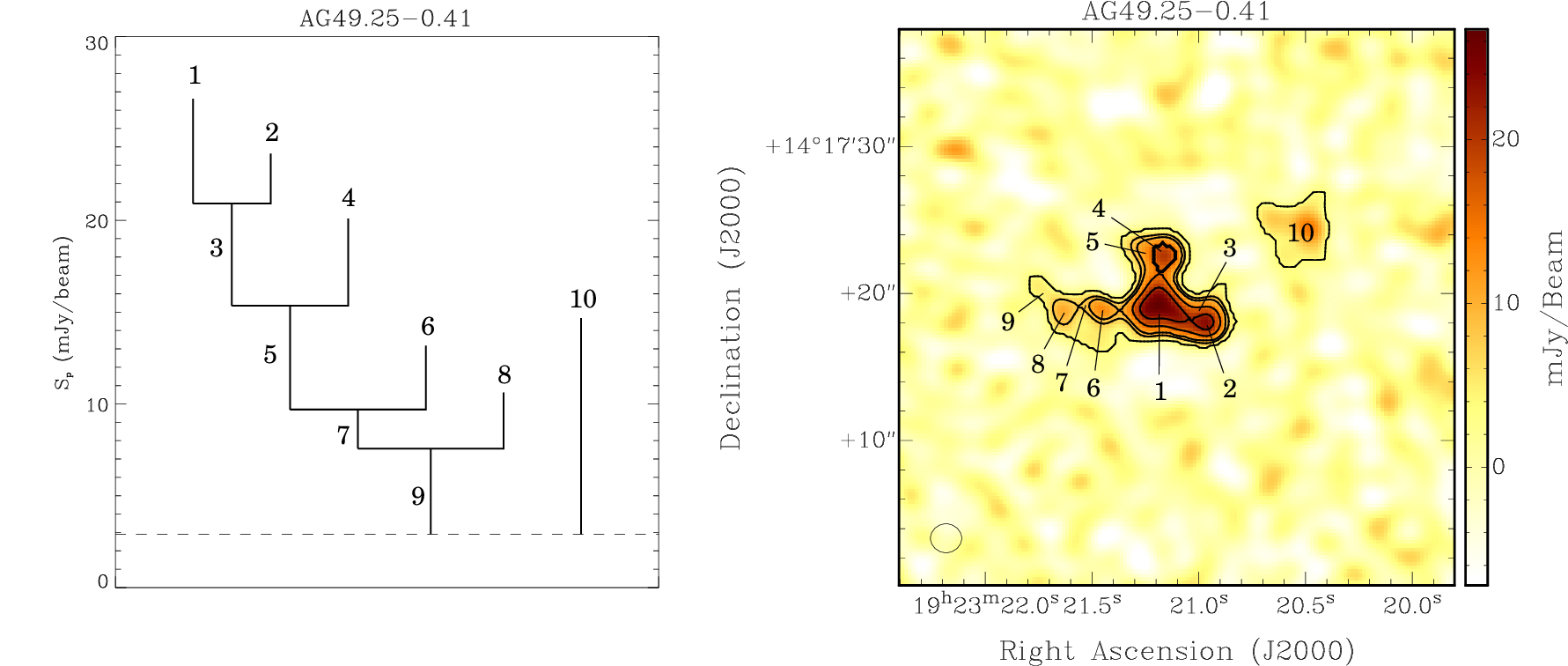}
	\caption{The left panel shows the dendrogram while the right panel shows the hierarchical structures in the source AG49.25$-$0.41. The dashed line in the left panel shows the $1\sigma$ noise. The contour levels in the right panel begin at the 1$\sigma$ level that defines the outer boundary of each root, with subsequent contours indicating the levels of nodes at which the root branches into two sub-structures. The synthesized beam is shown in the bottom left corner of the right panel. \label{dendro}}
\end{figure*}

We have carried out the dendrogram analysis for all the sources in our sample through the following approach. First, we determine local maxima that are significant at the 5$\sigma$ level. Next, contiguous regions were identified down to $\sim 1\sigma$ around each local maximum. Each distinct contiguous region was defined as a large scale ``core'', and corresponds to a \textit{root} in the dendrogram. All local maxima in a root that have a peak intensity that is at least 3$\sigma$, and are at least 1$\sigma$ above a node (i.e. the level at which local maxima merge into a larger structure), were defined as leaves. Figure~\ref{dendro} shows an example of the hierarchical structure in AG49.25$-$0.41 through a dendrogram.

While the dendrogram analysis is useful to examine the hierarchical structure in the clumps, one cannot directly use the analysis to determine the masses of the fragments. This is because the dendrogram only gives the height of a leaf above that of a node (i.e. the height of a peak above a saddle point). This by itself is not an estimate of the total flux density associated with the leaf. One would thus have to carry out an analysis using \clfind\ or \gclumps\ to determine the flux density and the mass associated with each fragment.

\begin{figure}[h!]
	\centering
	\includegraphics[width=0.5\textwidth]{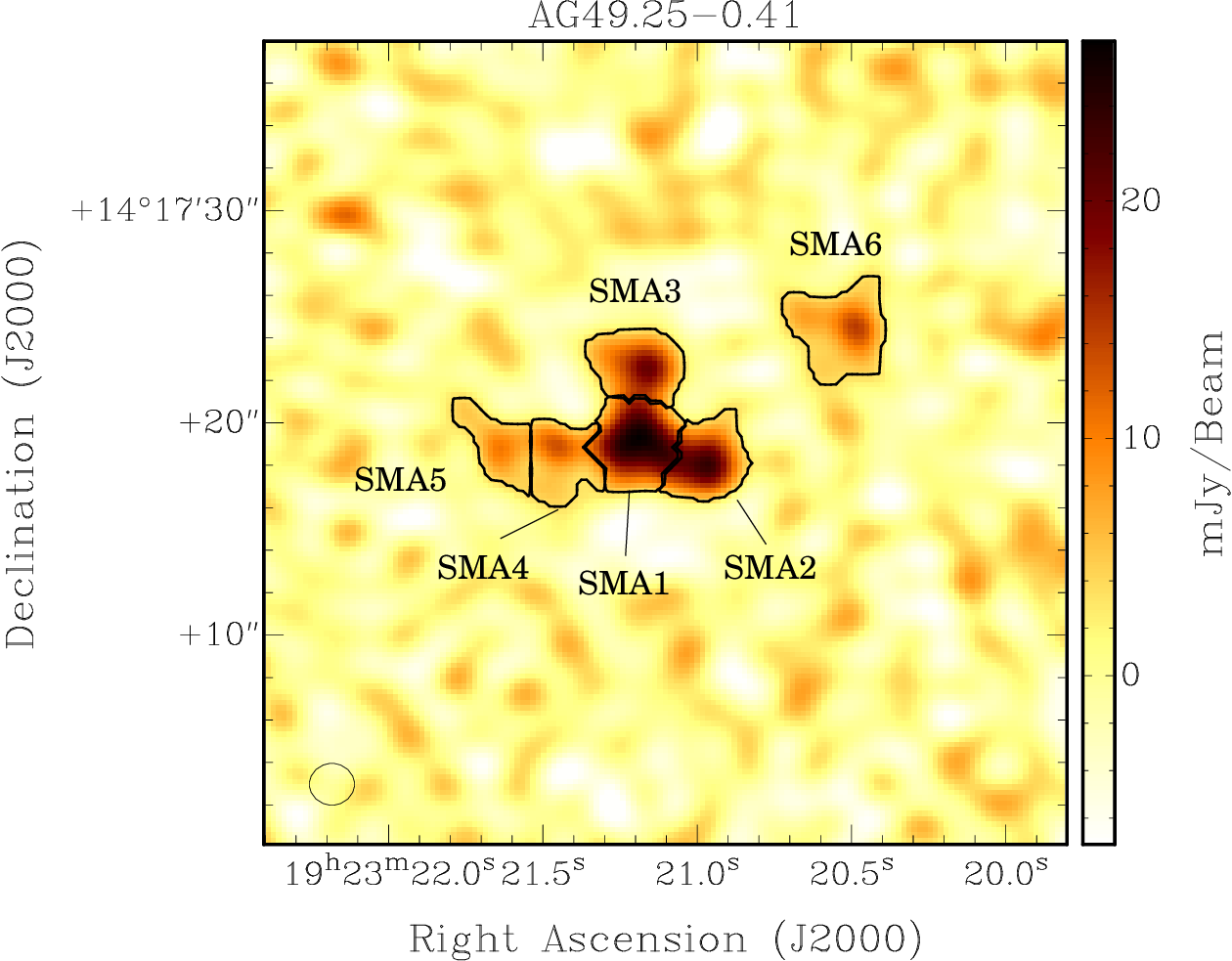}
	\caption{Fragments identified using \clfind\ for the source AG49.25$-$0.41. \label{clfind:example}}
\end{figure}

Both \clfind\ and \gclumps\ have merits and demerits with regards to determining the properties of the fragments. \clfind\ identifies isolated contours as individual clumps and blended contours that surround more than one local maximum are split using a ``friends-of-friends'' algorithm \citep{clumpfind}. In sources with several peaks and diffuse extended emission, this can give rise to fragments with odd shapes. An additional factor to be kept in mind is that the output from \clfind\ is at times a sensitive function of the choice of contour levels that are provided as an input to the algorithm. For our work, we mitigate this issue by using the flux densities of nodes that are identified in the dendrogram analysis as the input contour levels for \clfind. Since each node is a saddle point at which two local maxima merge together, the flux density of a node is the appropriate threshold contour level that can separate the two local maxima into individual fragments. This also ensures that spurious fragments are not identified due to closed contours that may arise due to noise in the image. Figure~\ref{clfind:example} shows the output of \clfind\ for the source AG49.25$-$0.41 where 6 distinct fragments have been identified, some with odd shapes. 

On the other hand, \gclumps\ fits the emission from the source using multiple Gaussian components. While this works well when the emission is comprised of several compact sources (which may be blended), difficulties are encountered when the overall morphology is complex with a mix of compact and extended emission. In such cases, some Gaussian components can be introduced purely for the sake of fitting the overall emission structure even though they may not be bona-fide fragments. This may in addition result in the flux densities of some fragments being under-estimated since part of the emission may be fit by an extended component. Considering the complex emission morphology in some sources, we choose to use \clfind\ in tandem with the dendrogram to identify fragments and determine their properties (see Table~\ref{fraglist}). It should be noted that both the dendrogram and \clfind\ analysis were first carried out using the CLEAN images which have uniform noise across the image, after which the source properties were extracted from the images that have been corrected for the primary beam response of the SMA antenna. 

The masses in Table~\ref{fraglist} have been calculated using the same temperature as for the ATLASGAL clump (Table~\ref{srclist}). However, it is likely that the temperature at the scales probed by the SMA is larger than what is observed at the clump scale of ATLASGAL. An extreme example is the case of the hot core G351.77$-$0.54 which was observed to have brightness temperatures close to 1000~K at a scale of $18-40$~AU \citep{beuther18}. Since our observations are at scales of $\sim 3000 - 20,000$~AU, we do not expect to see such high temperatures. Nonetheless, it is likely that the temperatures are higher than what is seen at 18\arcsec\ resolution, especially for sources with $\epsilon \gtrsim 1$. To explore this in more detail, we have looked at earlier studies of massive star forming regions at high resolution. \citet{wiseman98} studied the Orion ridge region using observations of ammonia with the Very Large Array (VLA) at a resolution of $8.5-9.0$\arcsec\ (which corresponds to physical scales comparable to that probed in our study since our sources are at much larger distances), and found temperatures to range from $\sim 15-30$~K in filaments although values as high as $\sim 80$~K are observed in the central core. \citet{issac20} studied the extended green object G19.88$-$0.53 using ALMA and found temperatures ranging from $47-116$~K, with high temperatures being characteristic of hot cores which correlates with the presence of an ionized radio jet in the region. On the other hand, \citet{wang12} studied the region G28.34+0.06 using ammonia observations with the VLA at $\sim 2$\arcsec\ resolution and found temperatures to range from $8-30$~K with higher temperatures coinciding with heating induced by outflows impinging on molecular gas. Based on this, we expect high temperatures in the sources AG48.61+0.02 and AG50.28$-$0.39, which have $\epsilon > 1$ and $L > 10^5~L_\sun$ with accompanying significant internal heating. Although these sources also have the highest dust temperatures at 18\arcsec\ resolution, the temperatures at high resolution may be significantly higher. An increase in temperature by a factor of 2 will reduce the masses of the cores by a similar factor. Thus, the masses of the massive fragments associated with high luminosity ATLASGAL clumps should be taken as upper limits, with the true masses being a factor of 1.5 to 2 smaller depending on the extent of central heating in the sources. However, considering the results of \citet{wiseman98, wang12}, we do not expect the masses of fragments that have no counterparts at mid-infrared wavelengths to be very different from what is reported in Table~\ref{fraglist}.

\begin{deluxetable*}{cccccccccccc}
\tablecaption{Fragments detected at 850~\micron\ using the compact configuration of SMA \label{fraglist}}
\tablewidth{0pt}
\tabletypesize{\scriptsize}
\tablehead{
	\colhead{Source name} & \colhead{$ \alpha $} & \colhead{$ \delta $} & \colhead{$ S_p $} & \colhead{$ S_\mathrm{int} $} & \colhead{$ M $} & \colhead{$ \Omega $} & \colhead{$R_\mathrm{eff}$} & \colhead{$ \Sigma $} & \colhead{70~\micron?} & \colhead{24~\micron?} & \colhead{MIR?} \\
	\colhead{} & \colhead{(J2000)} & \colhead{(J2000)} & \colhead{(mJy beam$^{-1}$)} & \colhead{(mJy)} & \colhead{($ M_\sun $)} & \colhead{(arcsec$^2$)} & \colhead{(pc)} & \colhead{(g cm$^{-2}$)} & \colhead{} & \colhead{} & \colhead{} 
}
\colnumbers
\startdata
AG36.79 SMA1 & $18^h59^m13^s.54$ & $03\degr20\arcmin36.8\arcsec$ & 28.6 & 94.8 & 74.7 & 17.3 & 0.10 & 0.50 & n & n & n \\
AG36.79 SMA2 & $18^h59^m12^s.52$ & $03\degr20\arcmin42.6\arcsec$ & 15.7 & 47.4 & 37.3 & 15.6 & 0.094 & 0.28 & n & n & n \\
AG36.79 SMA3 & $18^h59^m13^s.05$ & $03\degr20\arcmin35.6\arcsec$ &  9.7 & 23.7 & 18.7 & 12.1 & 0.083 & 0.18 & n & n & n \\
\hline
AG36.83 SMA1 & $18^h58^m41^s.03$ & $03\degr26\arcmin47.2\arcsec$ & 32.5 & 46.3 & 9.7 & 5.5 & 0.023 & 1.22 & n & n & n \\
AG36.83 SMA2 & $18^h58^m40^s.94$ & $03\degr26\arcmin50.9\arcsec$ & 29.5 & 42.2 & 8.9 & 5.6 & 0.023 & 1.10 & y & y & y \\
AG36.83 SMA3 & $18^h58^m41^s.23$ & $03\degr26\arcmin43.2\arcsec$ & 12.6 & 42.2 & 8.9 & 17.4 & 0.040 & 0.35 & n & n & n \\
\hline
AG36.90 SMA1 & $19^h00^m08^s.47$ & $03\degr20\arcmin32.0\arcsec$ & 132.8 & 382.7 & 217 & 14.7 & 0.090 &  1.78 & y & y & y \\
\hline
AG41.05 SMA1 & $19^h07^m12^s.64$ & $07\degr06\arcmin23.5\arcsec$ & 198.8 & 487.4 & 176 & 11.5 & 0.078 & 1.92 & y & y & n \\
\hline
AG41.07 SMA1 & $19^h06^m49^s.03$ & $07\degr11\arcmin07.1\arcsec$ & 40.8 & 71.2 & 27.7 & 6.8 & 0.062 & 0.49 & n & y & y \\
AG41.07 SMA2 & $19^h06^m49^s.10$ & $07\degr11\arcmin10.8\arcsec$ & 37.2 & 71.6 & 27.8 & 7.9 & 0.066 & 0.42 & y & n & y \\
AG41.07 SMA3 & $19^h06^m48^s.49$ & $07\degr11\arcmin12.3\arcsec$ & 21.0 & 83.1 & 32.3 & 18.3 & 0.10 & 0.21 & n & n & y \\
\hline
AG46.17 SMA1 & $19^h17^m50^s.18$ & $11\degr31\arcmin10.6\arcsec$ & 33.3 & 52.2 & 5.6 & 5.7 & 0.022 & 0.75 & n & n & n \\
AG46.17 SMA2 & $19^h17^m49^s.77$ & $11\degr31\arcmin05.8\arcsec$ & 29.6 & 83.1 & 8.9 & 12.4 & 0.033 & 0.55 & n & n & n \\
AG46.17 SMA3 & $19^h17^m49^s.12$ & $11\degr31\arcmin17.3\arcsec$ & 37.0 & 84.6 & 9.0 & 9.7 & 0.029 & 0.71 & n & n & n \\
%AG46.17 SMA4 & $19^h17^m49^s.46$ & $11\degr31\arcmin13.8\arcsec$ & 12.9 & 15.9 & 1.7 & 3.4 & 0.39 \\
\hline
AG46.43 SMA1 & $19^h17^m16^s.82$ & $11\degr52\arcmin30.2\arcsec$ & 64.1 & 129.6 & 18.8 & 8.3 & 0.028 & 1.56 & y & y & y \\
AG46.43 SMA2 & $19^h17^m16^s.96$ & $11\degr52\arcmin27.2\arcsec$ & 32.6 & 74.9 & 10.8 & 9.8 & 0.031 & 0.76 & n & n & n \\
AG46.43 SMA3 & $19^h17^m16^s.62$ & $11\degr52\arcmin30.7\arcsec$ & 31.7 & 59.9 & 8.7 & 7.6 & 0.027 & 0.79 & n & n & n \\
AG46.43 SMA4 & $19^h17^m16^s.45$ & $11\degr52\arcmin30.2\arcsec$ & 30.2 & 59.9 & 8.7 & 8.1 & 0.028 & 0.74 & n & n & n \\
\hline
AG47.03 SMA1 & $19^h16^m41^s.49$ & $12\degr38\arcmin11.3\arcsec$ & 74.5 & 143.2 & 46 & 7.8 & 0.057 & 0.95 & n & n & n \\
\hline
AG47.05 SMA1 & $19^h16^m41^s.42$ & $12\degr39\arcmin25.6\arcsec$ & 38.2 & 78.4 & 38 & 8.5 & 0.059 & 0.74 & n & n & n \\
\hline
AG48.61 SMA1 & $19^h20^m31^s.17$ & $13\degr55\arcmin25.6\arcsec$ & 1247.9 & 1880.9 & 393 & 5.1 & 0.060 & 7.2 & y & \textdagger & y\tablenotemark{a} \\
AG48.61 SMA2 & $19^h20^m30^s.83$ & $13\degr55\arcmin21.4\arcsec$ & 48.0 & 104.2 & 21.8 & 8.7 & 0.079 & 0.23 & n & \textdagger & n \\
\hline
AG49.25 SMA1 & $19^h23^m21^s.19$ & $14\degr17\arcmin19.3\arcsec$ & 27.6 & 55.8 & 13.1 & 8.3 & 0.043 & 0.48 & n & n & n \\
AG49.25 SMA2 & $19^h23^m20^s.96$ & $14\degr17\arcmin18.1\arcsec$ & 24.8 & 34.6 & 8.1 & 4.6 & 0.032 & 0.53 & n & n & n \\
AG49.25 SMA3 & $19^h23^m21^s.15$ & $14\degr17\arcmin22.6\arcsec$ & 20.2 & 29.4 & 6.9 & 5.0 & 0.033 & 0.42 & n & n & n \\
AG49.25 SMA4 & $19^h23^m21^s.44$ & $14\degr17\arcmin18.8\arcsec$ & 14.8 & 18.4 & 4.3 & 3.5 & 0.028 & 0.37 & n & n & n \\
AG49.25 SMA5 & $19^h23^m21^s.63$ & $14\degr17\arcmin18.8\arcsec$ & 13.1 & 15.1 & 3.5 & 2.7 & 0.024 & 0.39 & n & n & n \\
AG49.25 SMA6 & $19^h23^m20^s.48$ & $14\degr17\arcmin24.6\arcsec$ & 17.1 & 28.0 & 6.6 & 6.1 & 0.037 & 0.33 & n & n & n \\
\hline
AG49.60 SMA1 & $19^h23^m26^s.35$ & $14\degr40\arcmin18.2\arcsec$ & 145.3 & 671.3 & 124 & 18.9 & 0.064 & 1.99 & y & y & y \\
AG49.60 SMA2 & $19^h23^m26^s.87$ & $14\degr40\arcmin11.2\arcsec$ &  41.9 &  67.1 & 12.4 & 5.3 & 0.034 & 0.71 & n & n & n \\
\hline
AG50.28 SMA1 & $19^h25^m17^s.94$ & $15\degr12\arcmin24.7\arcsec$ & 564.2 & 1102.6 & 267 & 10.8 & 0.086 & 2.42 & y & \textdagger & y \\
\hline
AG52.85 SMA1 & $19^h31^m23^s.72$ & $17\degr19\arcmin46.8\arcsec$ & 331.3 & 1105.5 & 154 & 15.8 & 0.060 & 2.9 & y & n & n \\
AG52.85 SMA2 & $19^h31^m23^s.25$ & $17\degr19\arcmin48.0\arcsec$ & 159.7 & 321.0 & 45 & 8.7 & 0.044 &  1.53 & n & n & n \\
AG52.85 SMA3 & $19^h31^m24^s.40$ & $17\degr19\arcmin46.3\arcsec$ & 26.7 & 88.3 & 12.3 & 15.6 & 0.059 & 0.23 & n & y & y \\
\hline
AG52.92 SMA1 & $19^h27^m34^s.96$ & $17\degr54\arcmin38.1\arcsec$ & 228.8 & 266.8 & 84 & 2.8 & 0.029 & 6.8 & y & y$^\dagger$ & y \\
AG52.92 SMA2 & $19^h27^m35^s.17$ & $17\degr54\arcmin46.9\arcsec$ & 50.3 & 122.0 & 38 & 10.3 & 0.055 & 0.85 & n & n & n \\
\hline
AG53.14 SMA1 & $19^h29^m17^s.49$ & $17\degr56\arcmin18.7\arcsec$ & 671.1 & 1445.7 & 19 & 11.1 & 0.014 & 6.1 & n & n & n \\
AG53.14 SMA2 & $19^h29^m17^s.58$ & $17\degr56\arcmin23.4\arcsec$ & 447.6 &  722.9 & 9.5 & 7.4 & 0.012 & 4.6 & y & y$^\dagger$ & y \\
AG53.14 SMA3 & $19^h29^m18^s.17$ & $17\degr56\arcmin14.9\arcsec$ &  57.2 &  133.3 & 1.8 & 12.2 & 0.015 & 0.51 & n & n & n \\
AG53.14 SMA4 & $19^h29^m18^s.00$ & $17\degr56\arcmin15.7\arcsec$ &  47.0 &  116.4 & 1.5 & 13.1 & 0.016 & 0.41 & n & n & n \\
AG53.14 SMA5 & $19^h29^m17^s.00$ & $17\degr56\arcmin25.9\arcsec$ &  43.0 &   80.0 & 1.1 & 9.1 & 0.013 & 0.41 & n & n & n \\
AG53.14 SMA6 & $19^h29^m16^s.83$ & $17\degr56\arcmin23.9\arcsec$ &  37.5 &   57.6 & 0.76 & 6.7 & 0.011 & 0.40 & n & n & n \\
AG53.14 SMA7 & $19^h29^m16^s.53$ & $17\degr56\arcmin27.4\arcsec$ &  37.7 &   53.9 & 0.71 & 5.9 & 0.011 & 0.42 & n & n & n \\
\enddata
\tablecomments{The columns show (1) the source name (2)$-$(3) J2000 equatorial coordinates (4) peak flux density (primary beam corrected) (5) integrated flux density (primary beam corrected) (5) mass (6) deconvolved solid angle (7) effective radius and (8) mass surface density. Columns (9), (10) and (11) indicate the presence or absence of 70~\micron\ counterpart in the Hi-GAL survey, 24~\micron\ counterpart in the MIPSGAL survey (\textdagger\ indicates saturation) and an MIR counterpart in the GLIMPSE survey respectively. \tablenotetext{a}{Not a point source; possible Extended Green Object (EGO)}}
 \end{deluxetable*}

\subsection{Notes on selected sources}\label{sources:individual}

\begin{figure*}[h]
	\centering
	\includegraphics[width=\textwidth]{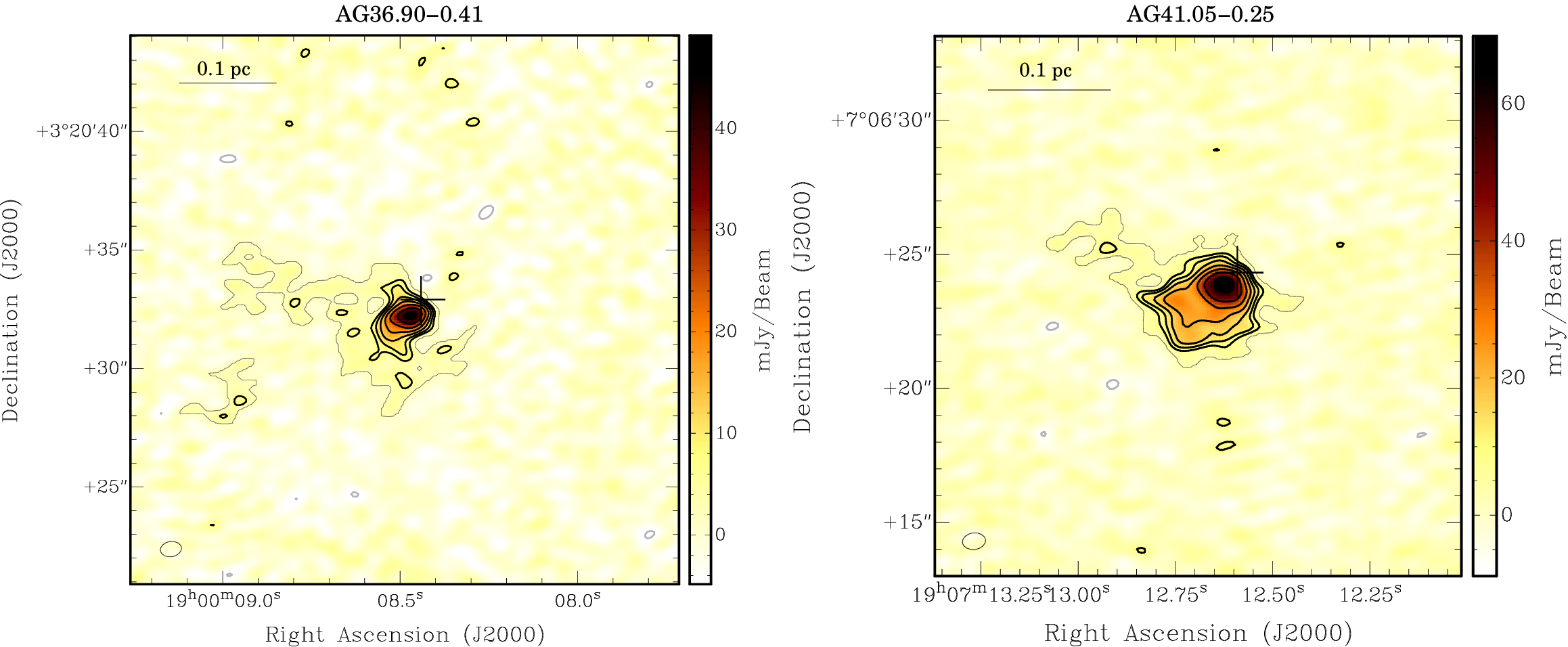}
	\caption{The left and right panels show AG36.90$-$0.41 and AG41.05$-$0.25 respectively as seen by the SMA at 0.6\arcsec\ resolution. The images have been made by combining the data from the compact and extended configurations of the SMA. The grey contours show emission at the $-3\sigma$ level. The thin black contour is at $1\sigma$, while the thick black contours start at $3\sigma$ and increase thereafter by factors of $\sqrt{2}$. The synthesized beam is shown in the bottom left. \label{ag3690:ag4105}}
\end{figure*}

\textit{AG36.90$-$0.41 and AG41.05$-$0.25} -- Both sources are dominated by a single structure at $\sim 2\arcsec$ resolution, although some extended emission is seen surrounding the central source. Figure~\ref{ag3690:ag4105} shows the sources at 0.6\arcsec\ resolution by combining the compact and extended configurations of the SMA. Clearly, the emission continues to be dominated by a single core even at high resolution. Both sources are associated with a 24~\micron\ point source with AG36.90$-$0.41 also hosting a 6.7~GHz methanol maser \citep{pandian11}.

\textit{AG48.61+0.02} -- There are two fragments in this source, with SMA1 dominating the emission with a mass of 393~$M_\sun$ (although the true mass may be much lower depending on the temperature of the core at this scale). The other fragment, SMA2 has a mass of only 21.8~$M_\sun$. The infrared emission from the clump is also very strong with the entire clump being saturated in the MIPSGAL survey. The morphology of emission in the GLIMPSE survey is suggestive of there being an outflow along with extended green emission (i.e. excess emission at 4.5~\micron).

\textit{AG53.14+0.07} -- There are seven fragments detected in this clump with masses ranging from 0.7~$M_\sun$ to 19~$M_\sun$. The clump is also associated with a 6.7~GHz methanol maser whose location is closest to that of SMA2 \citep{pandian11}. SMA2 is also associated with a point source in both GLIMPSE and MIPSGAL surveys although no point source is listed in the catalogs due to possible saturation.

\section{Discussion}\label{section:discussion}
\subsection{Nature of fragments}\label{frag:nature}

We have detected a total of 43 fragments in 16 clumps. We have searched for counterparts at mid-infrared wavelengths at wavelengths shorter than 70~\micron\ using the Hi-GAL, MIPSGAL and GLIMPSE catalogs using a search radius of 3\arcsec\ (see Table~\ref{fraglist}). We find that 8 fragments have point source counterparts in the GLIMPSE catalog. An additional three fragments have emission from a point source in the GLIMPSE maps although they are not listed in the catalog. The infrared colors in the Spitzer-IRAC bands are consistent with 5 out of 8 fragments being Class I young stellar objects (YSOs) and one source (AG410.7$-$0.12 SMA3) being a Class II YSO \citep{gutermuth08}. The nature of the two other GLIMPSE sources is not clear since they are detected in only two IRAC bands.

Seven fragments have a counterpart in the MIPSGAL catalog, with an additional two sources having a point source that is saturated in the central region. We also find that the 24~\micron\ emission in MIPSGAL is fully saturated around AG48.61+0.02 and AG50.25$-$0.39. Eleven fragments have associated 70~\micron\ counterparts in the Hi-GAL survey. Thus, around 28 fragments have no counterpart at mid-infrared wavelengths and may be starless, although further studies using deeper observations are required to establish the same. Of these, AG36.79~SMA1, AG36.79~SMA2, AG47.03~SMA1, AG47.05~SMA1, AG52.82~SMA2 and AG52.92~SMA2 have masses exceeding 20~$M_\sun$, and may be suitable candidates for being massive starless cores.

\subsection{Fragmentation of the ATLASGAL clumps}\label{frag:properties}

The fragmentation properties and the underlying processes that lead to fragmentation is one of the outstanding questions in the study of massive star formation. According to the Jeans instability criterion, fragmentation occurs when the mass exceeds the Jeans mass ($M_J$) over a Jeans length ($\lambda_J$). The Jeans length for thermal fragmentation is given by

\begin{equation}
\lambda_J = \sqrt{\frac{\pi c_s^2}{G\langle \rho \rangle}}
\end{equation}
where $c_s^2 = kT/\mu m_H$ is the sound speed and $\langle \rho \rangle$ is the mean density in the clump. The Jeans mass is given by
\begin{equation}
M_J = \frac{4}{3}\pi \left( \frac{\lambda_J}{2} \right)^3 \langle \rho \rangle
\end{equation}
The Jeans length and Jeans mass for fragmentation that is governed by turbulence rather than thermal motion can be derived by replacing the isothermal sound speed with the turbulent velocity dispersion
\begin{equation}
\sigma_\mathrm{turb}^2 = \sigma_\mathrm{CO}^2 - \frac{kT}{\mu_\mathrm{CO}m_H}
\end{equation}

\begin{figure*}[h!]
	\centering
	\includegraphics[width=0.9\textwidth]{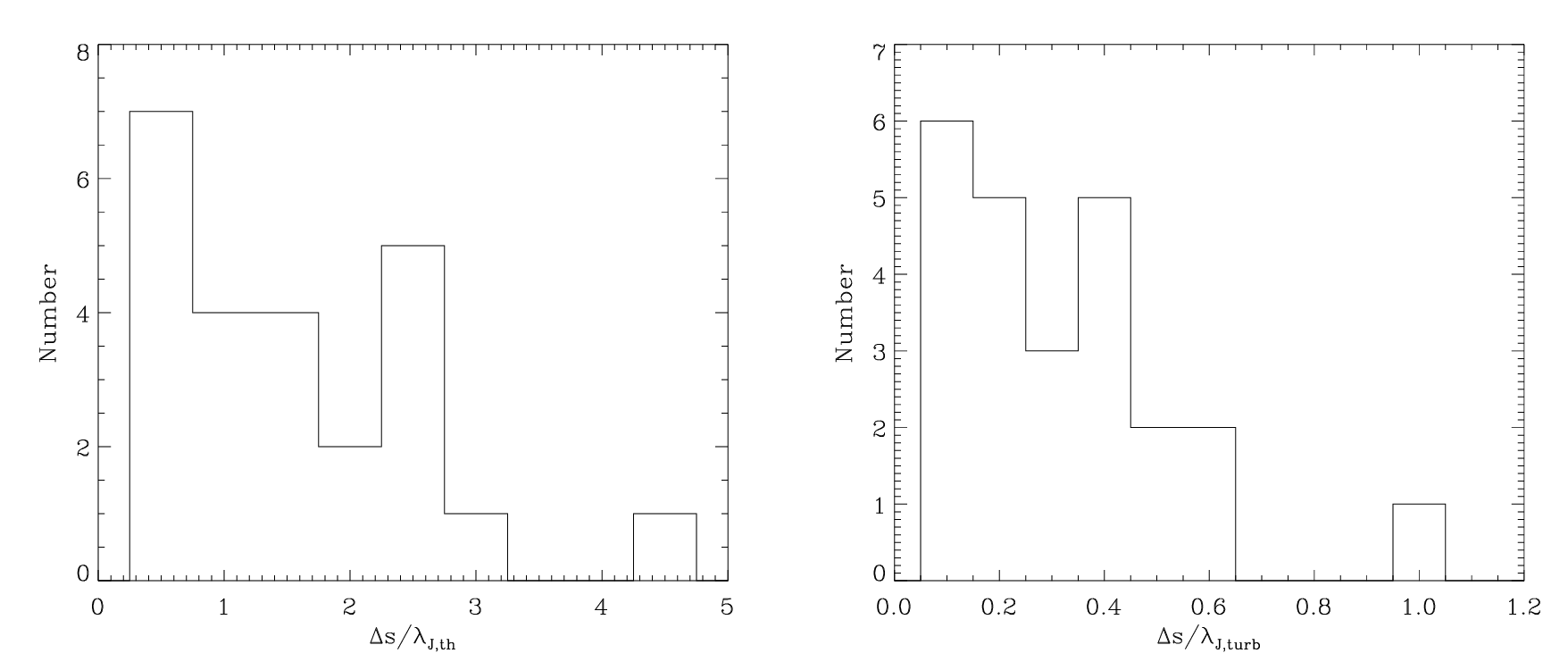}
	\caption{The ratio of the nearest neighbor separations between fragments and the thermal Jeans length (left panel) and turbulent Jeans length (right panel). \label{jeanslength}}
\end{figure*}

Table~\ref{fragprop} shows the thermal and turbulent Jeans length and Jeans mass for our sample along with the range of nearest neighbor separations for the fragments detected with the SMA. It is to be noted that these separations are observed in projection and are thus lower limits of the true inter-fragment separation. We have used the dust temperature (see Table~\ref{srclist}) to estimate the thermal sound speed, while the mean density has been computed from the clump mass and effective radius using 
\begin{equation}
\langle \rho \rangle = \frac{3M}{4\pi R_\mathrm{eff}^3}
\end{equation}
Figure~\ref{jeanslength} shows a histogram of the ratio of the nearest neighbor separations to the thermal and turbulent Jeans length.

\begin{deluxetable*}{cccccccc}
\tablecaption{Fragmentation properties of the ATLASGAL clumps \label{fragprop}}
\tablewidth{0pt}
\tablehead{
	\colhead{Source name} & \colhead{$\lambda_{J,th}$} & \colhead{$\lambda_{J,turb}$} & \colhead{$\Delta s$} & \colhead{$M_{J,th}$} &  \colhead{$M_{J,turb}$} & \colhead{$M_\mathrm{frag}$} & \colhead{$M_{5\sigma}$} \\
	\colhead{} & \colhead{(pc)} & \colhead{(pc)} & \colhead{(pc)} & \colhead{($M_\sun$)} & \colhead{($M_\sun$)} & \colhead{($M_\sun$)} & \colhead{($M_\sun$)}
	}
\colnumbers
\startdata
AG36.79$-$0.20 & 0.19 & 1.4  & $0.31 - 0.45$ & 3.4 & 1365 & $19 - 74$   & 7.0 \\
AG36.83$-$0.04 & 0.06 & 0.37 & $0.07 - 0.09$ & 0.8 & 215  & $8.9 - 9.8$ & 2.5 \\
AG36.90$-$0.41 & 0.17 & 1.1  & \ldots        & 3.7 & 1038 & 217         & 8.7 \\
AG41.05$-$0.25 & 0.21 & 0.95 & \ldots        & 5.9 & 565  & 176         & 5.5 \\
AG41.07$-$0.12 & 0.31 & 1.5  & $0.16 - 0.38$ & 8.6 & 980  & $28 - 32$   & 4.8 \\
AG46.17$-$0.53 & 0.05 & 0.24 & $0.13 - 0.25$ & 1.0 & 94   & $5.6 - 9.0$ & 1.4 \\
AG46.43$-$0.24 & 0.11 & 0.54 & $0.05 - 0.06$ & 1.8 & 238  & $8.7 - 19$  & 2.4 \\
AG47.03+0.24   & 0.12 & 1.4  & \ldots        & 5.7 & 1417 & 46          & 5.5 \\
AG47.05+0.25   & 0.11 & 0.77 & \ldots        & 2.2 & 690  & 38          & 9.6 \\
AG48.61+0.02   & 0.12 & 0.51 & 0.31          & 6.7 & 463  & $22 - 393$  & 6.4 \\
AG49.25$-$0.41 & 0.13 & 0.59 & $0.07 - 0.25$ & 2.8 & 249  & $3.5 - 13$  & 3.4 \\
AG49.60$-$0.25 & 0.11 & 0.64 & 0.27          & 2.6 & 590  & $12 - 124$  & 6.0 \\
AG50.28$-$0.39 & 0.13 & 0.46 & \ldots        & 5.8 & 288  & 267         & 6.9 \\
AG52.85$-$0.66 & 0.09 & 0.44 & $0.18 - 0.26$ & 2.7 & 321  & $12 - 154$  & 3.3 \\
AG52.92+0.41   & 0.10 & 0.68 & 0.28          & 2.1 & 644  & $38 - 84$   & 7.0 \\
AG53.14+0.07   & 0.02 & 0.09 & $0.02 - 0.04$ & 0.7 & 34   & $0.7 - 19$  & 0.45 \\
\enddata
\tablecomments{The columns show (1) the source name (2) thermal Jeans length  (3) turbulent Jeans length (4) range of nearest neighbor separation (5) thermal Jeans mass (6) turbulent Jeans mass (7) range of fragment masses and (8) 5$\sigma$ mass sensitivity of the SMA observation}
\end{deluxetable*}

Table~\ref{fragprop} and Figure~\ref{jeanslength} show that the separation and masses of the fragments detected in our study are more consistent with thermal rather than turbulent fragmentation. The masses of all fragments are smaller than the turbulent Jeans mass, and the inter-fragment separations are much smaller than the turbulent Jeans length except for one source. Figure~\ref{jeanslength} shows that there are cases where the nearest neighbor separation is also smaller than the thermal Jeans length, similar to what was observed by \citet{beuther18}. However, all these cases pertain to fragments that are part of a larger structure, but are identified as a distinct leaf in the dendrogram. The sub-Jeans scale separations in these cases are likely to be a consequence of the local clump densities being higher than the mean density used to compute the thermal Jeans length. Though less likely, it is also possible that some fragments are extended structures pertaining to the main fragment rather than being independent fragments themselves. The results obtained here are in agreement with several statistical and detailed studies to date (e.g. \citealt{palau15, beuther18, palau18, beuther19, sanhueza19, svoboda19, saha22}) where the fragmentation has been found to be governed by thermal rather than turbulent motion. 

One can also observe significant diversity in the fragmentation properties of the sample. Five clumps are dominated by a single massive core with masses ranging from 38 to 276~$M_\sun$ with no low-mass cores to the sensitivity limit. Of these, the observations are sensitive to the thermal Jeans mass at the $5\sigma$ limit for three sources suggesting a real paucity of low-mass cores. Further, two sources continue to be dominated by a single fragment down to 0.6\arcsec\ resolution ($\sim 0.025$~pc scale). Another four clumps have the lowest fragment mass to be greater than 5~$M_{J,th}$, with the number increasing to nine clumps at the limit of 3~$M_{J,th}$, although one has to bear in mind the caveat that the observations are not sensitive to the thermal Jeans mass limit in some cases. This is similar to the observations of \citet{zhang09, bontemps10, palau13, zhang15, csengeri17, figueira18} where limited fragmentation was seen with the most massive fragment having masses several times that of the thermal Jeans mass. On the other hand, two clumps display significant fragmentation with the lowest fragment mass comparable to the thermal Jeans mass. This is similar to what was observed by \citet{cyganowski17, henshaw17, beuther18, zhang21} where a significant population of low-mass fragments were found to form coevally with massive fragments. We also do not see any correlation between the degree of fragmentation, quantified by the number of clumps, and the mean density. Since the Jeans mass decreases with increasing density, one expects to observe a higher degree of fragmentation with increasing density. Thus, although the fragment separations are closer to that expected from thermal fragmentation, there are other factors that control the degree of fragmentation. As pointed out by \citet{palau14, beuther18}, the diversity in fragmentation may be due to the density profile of the parent clump, with steeper density profiles seen to limit fragmentation. We plan to carry out a study of the density profiles of our ATLASGAL sample to verify whether this can explain our observed fragmentation diversity. Alternately, magnetic fields may also play an important role in the degree of fragmentation \citep{commercon11, fontani18}. For example, radiation magnetohydrodynamic simulations of \citet{commercon11} are able to inhibit fragmentation in highly magnetized cores, with the level of fragmentation dependent on the strength of the magnetic field \citep{fontani18}.

\subsection{The initial fragmentation of massive clumps}\label{frag:initial}

One of the primary differences between the theories of core accretion and competitive accretion is regarding the initial fragmentation of massive clumps. While the theory of core accretion predicts initial fragmentation into a population of cores whose mass function mimics that of the stellar initial mass function, the competitive accretion scenario has the clump fragmenting into a large number of fragments with mass comparable to the thermal Jeans mass. Observational studies obtain a snapshot of the state of fragmentation well after the initial stages, with it being difficult to establish the period of time that has elapsed since the beginning of fragmentation to the time of observation. It is hence not straightforward to make interpretations about the initial conditions of fragmentation based on observational results.

However, the observation of sources that are dominated by a single core (e.g. CygX-N63 of \citealt{bontemps10}, NGC~7129-FIRS2 and I20126+4104 of \citealt{palau13}, several ATLASGAL clumps in our work, and other works cited in the references above), and the paucity of low-mass fragments in some clumps (e.g. \citealt{zhang15}) suggests that the parent clumps in these cases did not fragment into a large number of Jeans mass fragments. This does not mean that massive stars form in isolation in these sources since the core mass is typically a small fraction of the clump mass, and low-mass stars may form in the clump at later times \citep{zhang15}.

\begin{figure*}[h]
\plotone{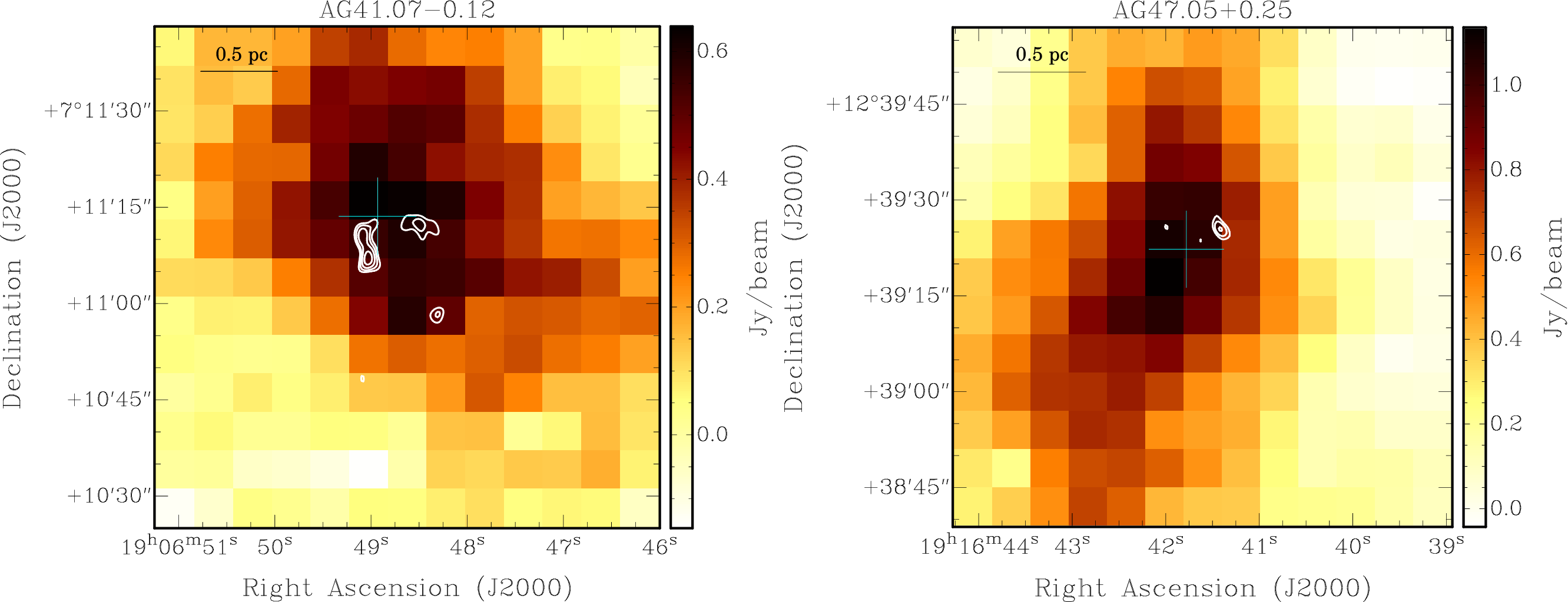}
\caption{The left and right panels show the continuum measured by ATLASGAL towards AG41.07$-$0.12 and AG47.05+0.25 with white contours depicting emission detected by the SMA at 2$''$ resolution. The position of the ATLASGAL clump is marked by the cyan cross. \label{ag4107-ag4705}}
\end{figure*}

An interesting observation in the context of initial fragmentation is the offset between the locations of the SMA fragments and that of the ATLASGAL clump. We see four sources (AG36.83$-$0.04, AG41.07$-$0.12, AG47.05+0.25 and AG52.92+0.41) where the offset exceeds 5\arcsec. Figure~\ref{ag4107-ag4705} shows the offsets for AG41.07$-$0.12 and AG47.05+0.25 respectively, with all three fragments in the former offset from the ATLASGAL source in excess of 13\arcsec, whereas the latter has a single fragment that is offset from the mean clump position by over 6\arcsec. A caveat in the context of this discussion is that the position of the ATLASGAL clump depends on the algorithm used to define the source. For example, the positions of ATLASGAL sources differ by several arcseconds between the compact source catalog \citep{atlasgalcsc} and the catalog generated by \gclumps\ \citep{csengeri14}. However, based on visual inspection, the positions of the SMA fragments are significantly offset from the mean clump position in all the cases above. Thus, the offset between the locations of the fragments and the large scale clump is real, subject to the astrometric accuracy of the two data sets.

The astrometry of the ATLASGAL data have been found to be accurate to that of the pointing of the APEX telescope which is $\sim 2-3''$. The astrometric accuracy of the SMA data is determined by the uncertainty in the positions of the gain calibrators, separation between the calibrators and target sources, uncertainty in the antenna positions and the signal to noise ratio of the target source itself. The positions of the gain calibrators are from VLBA calibrator surveys which have position acccuracies better than 1 milli-arcsecond (mas), while antenna positions are known to 0.1$-$0.2$\lambda$ (where $\lambda$ is the wavelength). As shown in Table~\ref{smaobs}, the SMA observations included two gain calibrators on all days of observations. For example, the 2010 October 4 data included 1751+096 and 2025+337 as gain calibrators, which are located about 26$^\circ$ from G47.05+0.25 respectively. The positions of 1751+096 and 2025+337 are known to better than 0.2 mas \citep{johnston95} and 1.1 mas \citep{beasley02} respectively. The uncertainty in astrometry from the phase transfer from gain calibrators to target sources can be estimated by examining the phase solutions over a time duration corresponding to the separation between the calibrators and targets (1.3 hours for 20$^\circ$ separation). An examination of the gain solutions shows that the astrometric uncertainty from this process is smaller than the size of the synthesized beam ($\sim 2''$). To further verify the accuracy of astrometry, we calibrated the data using only one gain calibrator, 1751+096, and imaged the second gain calibrator, 2025+337, which is almost 43$^\circ$ away. The position of 2025+337 agreed with the value given in the VLBA calibrator survey to within 0.5$''$. Since our target sources are much closer to the gain calibrator compared to the separation between the two gain calibrators above, we infer that the astrometric accuracy of the SMA fragments are better than 0.5$''$. Thus, the detected offsets between the positions of the ATLASGAL and SMA sources is larger than the astrometric uncertainty of the source positions and is clearly significant.

Finally, one has to consider whether the offsets are an outcome of differences in resolution. However, a difference in resolution by itself cannot introduce an offset between a compact structure (that is unresolved at poorer angular resolution) and the overall large scale structure. Alternately, a smoothing operation of an SMA dataset to the lower resolution of ATLASGAL cannot by itself change the location of the peak emission unless there are multiple compact sources of comparable strength present, in which case the smoothed data would show a peak at the mean position of the multiple sources. In cases where there is a single compact source, or where one source is significantly stronger than other sources in the vicinity, the smoothed data must have a peak emission that is identical to or close to the location of the strong compact source. Thus, the detected offsets between the ATLASGAL emission and our SMA maps are unlikely to be due to the effect of resolution alone. Considering that observations using the compact configuration of the SMA filter out emission on scales larger than $\sim 13$\arcsec\ (see Section~\ref{section:obs}), the observed offsets are most likely depicting the actual differences between the location of smoother diffuse clump emission that is measured by ATLASGAL and the compact cores that are discernible upon filtering of this extended emission by the interferometer (for example, in the case of AG47.05+0.25, over 90\% of the emission is filtered out by the SMA).

The angular offsets that exceed 5\arcsec\ correspond to a physical projected separations that are a significant fraction of the effective radius of the clump (for example, 0.57~pc in the case of AG41.07$-$0.12 and 0.22~pc in the case of AG47.05+0.25). If the process of forming a massive core involves competitive accretion of the gas reservoir onto a large number of Jeans mass fragments, the massive core must be located near the center of the gravitational potential \citep{bonnell01}. Since fragments compete among themselves to accrete the gas that is distributed within the clump, fragments that are far from the center of the potential do not accrete significantly and retain low masses \citep{bonnell06}. The observation of sub-millimeter fragments that are significantly offset from the center of the gravitational potential of the clump in at least four sources strongly suggests that the cores did not form from competitive accretion of the clump material onto a large number of fragments throughout the clump. We stress that these results primarily suggest that a clump does not undergo initial fragmentation into a large number of Jeans mass fragments as invoked by models of competitive accretion. It does not rule out the phenomenon of competitive accretion to transfer mass from clumps to multiple cores within that may have formed through limited fragmentation.

\subsection{Mass assembly from clumps to cores}\label{mass:assembly}

The ``core formation efficiency'' (CFE) is defined by \citet{bontemps10, palau13, csengeri17} as the fraction of the clump mass that is in compact structures. We have estimated the CFE for each source by taking the ratio of the total flux density of all fragments detected in the SMA map to that in the ATLASGAL map. The latter is computed by taking the total flux density that is within the 10\% level of the SMA primary beam centered at the SMA pointing location. It is important to note that estimating the CFE as the ratio of flux densities  is based on the assumption that the temperatures at high and low resolution are identical. As noted in Section~\ref{section:results}, the actual temperature at high resolution may be significantly higher, especially in sources with $\epsilon > 1$. This results in the CFE to be overestimated in such sources. To account for this effect, we incorporate an additional uncertainty in temperature by a factor of 2 for sources with $\epsilon > 1$ and a factor of 1.5 for sources with $\epsilon > 0.1$ for estimating the CFE.

\begin{figure*}[h]
	\centering
	\includegraphics[width=0.9\textwidth]{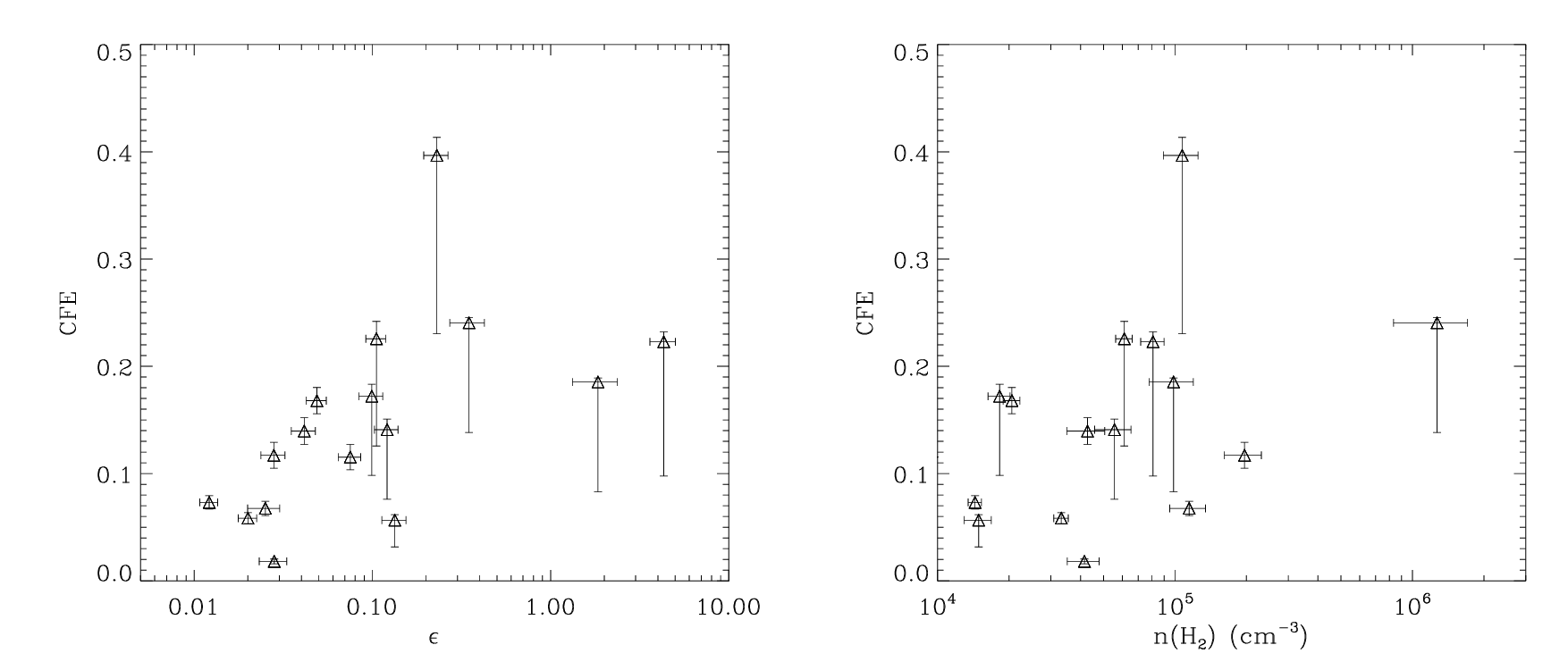}
	\caption{The core formation efficiency (CFE) as a function of (a) the evolutionary state parameter (left panel) and (b) mean H$_2$ number density (right panel). \label{cfe}}
\end{figure*}

Figure~\ref{cfe} shows the CFE as a function of $\epsilon$ (eq.~\ref{eq:mlratio}), the evolutionary state parameter (left panel) and the mean number density of H$_2$ (right panel) in the ATLASGAL clump. We observe a positive correlation between the CFE and $\epsilon$, with a Spearman correlation coefficient of 0.69 and a $p-$value of 0.003. This shows that the two quantities are correlated with only a 0.3\% probability of the correlation occurring by chance. It is to be noted that the correlation coefficient does not take into account the uncertainties associated with the two quantities. However, as seen from Figure~\ref{cfe}, accounting for the uncertainty in CFE only reduces the correlation coefficient and doesn't remove it altogether. In fact, considering the CFE to be at the lower limits implied by the uncertainty only reduces the correlation coefficient to 0.31. Thus, we find that compact structures encompass a larger fraction of the clump mass as the source evolves. This in turn shows that mass is channeled from clumps to cores over time and strongly suggests that the evolution of a core does not happen in isolation as in models of core accretion (e.g. \citealt{krumholz09}). We also observe a weak correlation between the CFE and the density of the ATLASGAL clump, with a Spearman correlation coefficient of 0.49 and a $p-$value of 0.06. This is similar to the observations of \citet{bontemps10, palau13, csengeri17}, although our data does not adequately sample H$_2$ number densities greater than $10^6$~cm$^{-3}$.

\begin{figure*}[h]
	\centering
	\includegraphics[width=0.9\textwidth]{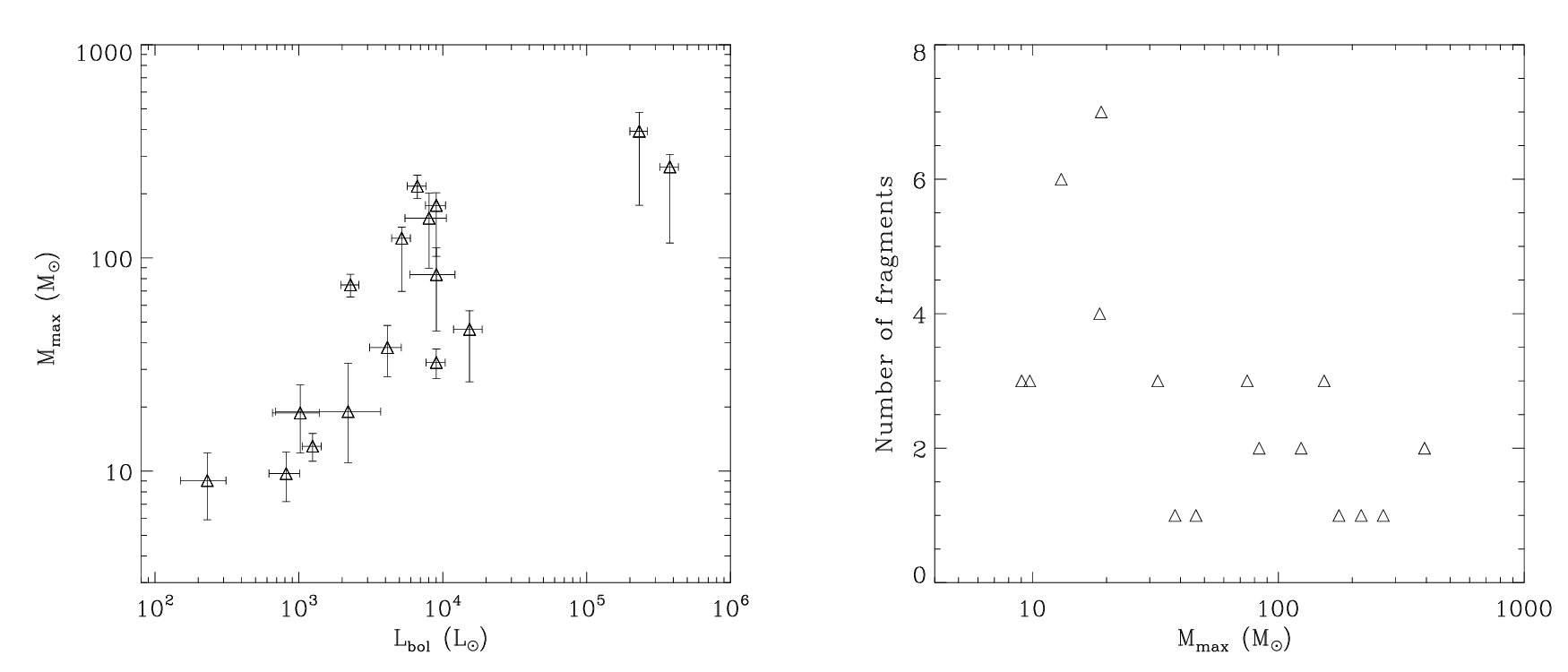}
	\caption{The left panel shows the maximum mass of a fragment as a function of bolometric luminosity of the clump. The right panel shows the number of fragments as a function of the maximum fragment mass in a clump. \label{maxmass:plots}}
\end{figure*}

We also find a good correlation between the maximum mass of a fragment ($M_\mathrm{max}$) in a clump and the bolometric clump luminosity (left panel of Figure~\ref{maxmass:plots}, with the uncertainty in maximum mass incorporating the heating effect as outlined in the previous paragraph). A Spearman correlation test gives a correlation coefficient of 0.81 with a $p-$value of $1.4 \times 10^{-4}$. Since both the mass and luminosity are dependent on the distance, we also carried out a partial Spearman correlation test to remove the effect of distance. This gives a correlation coefficient of 0.6 which indicates that the two quantities are well correlated. The degree of correlation seen in our work is much stronger than that of \citet{palau13} who observed only a weak correlation in their sample. As indicated by \citet{palau13}, the presence of this correlation argues against a formation scenario purely by competitive accretion since such a mechanism precludes any correlation between the mass of small scale condensations and the final stellar mass \citep{bonnell04}. We also do not see any positive correlation between the maximum mass of a fragment and the number of fragments (right panel of Fig~\ref{maxmass:plots}). The correlation is in fact negative with a correlation coefficient of $-0.65$ with a $p-$value of 0.006, whereas \citet{bonnell04} predict an increase in the number of stars with increasing maximum mass. Since competitive accretion models predict the number of stars to increase with the maximum mass in the cluster, the lack of a positive correlation between the number of fragments and the maximum mass of a fragment also suggests that massive stars do not form purely by competitive accretion.

The transfer of mass from clump to core scales with evolution coupled with evidence against pure competitive accretion suggests that massive stars grow through clump fed core accretion (e.g. \citealt{wang10}). This is also corroborated by observations such as that of \citet{zhang11, yuan18, peretto20} and references therein. Competitive accretion is likely to play a role in the process of mass transfer from clump to core scales when multiple cores are present in the clump.

\section{Conclusions}\label{section:conclusion}

We have carried out a study of fragmentation in massive clumps using observations of ATLASGAL sources with the SMA in its compact configuration at an angular resolution of 2\arcsec. Our main conclusions are
\begin{enumerate}
\item We observe a wide diversity of fragmentation in the sample. Some clumps are dominated by a single core or have limited fragmentation with the least massive fragment having a mass several times the Jeans mass, while other clumps show significant fragmentation with masses down to the Jeans mass. Two sources continue to be dominated by a single core even when observed with the extended configuration of the SMA at a resolution of 0.6\arcsec, or a physical scale of $\sim 0.025$~pc. 
\item The separation and masses of the fragments are smaller than the Jeans length and Jeans mass inferred from the turbulent velocity dispersion. Thus, the data are more consistent with thermal rather than turbulent fragmentation.
\item Four sources show a significant projected separation between the clump and the cores within, which taken together with the observation of limited fragmentation in many sources suggests that clumps do not undergo initial fragmentation into a large number of thermal Jeans mass fragments.
\item We find evidence for a larger fraction of clump mass to be incorporated into cores with evolution. 
\item We also find a good correlation between the mass of the most massive fragment in the clump and the bolometric luminosity. We do not see the number of fragments to increase with increasing mass of the most massive fragment. These results suggest that massive stars do not form purely by competitive accretion.
\end{enumerate}

Our results support a broad picture of massive star formation wherein the initial fragmentation of massive clumps is governed by thermal fragmentation in combination with the underlying density distribution and/or magnetic fields. Subsequently, cores continue to gain mass from the clump, a process that may incorporate aspects of competitive accretion depending on the number of fragments within the clump. Individual massive stars or binary systems are likely form within these cores, as seen in high resolution observations of \citet{beuther17, motogi19}.

\section*{Acknowledgements}
We thank the referee for valuable comments that helped improve the manuscript. J.D.P. thanks the Max Planck Society for funding this research through the Max Planck Partner Group. The Submillimeter Array is a joint project between the Smithsonian Astrophysical Observatory and the Academia Sinica Institute of Astronomy and Astrophysics and is funded by the Smithsonian Institution and the Academia Sinica. This research has utilized NASA's Astrophysics Data System, CDS's VizieR catalog access tool, and calibrated infrared images from the Spitzer Heritage Archive and Herschel Science Archive.

\vspace{5mm}
\facilities{SMA, APEX, Herschel, Spitzer}

\bibliography{hmsf} 
\bibliographystyle{aasjournal}

\appendix
\section{Source photometry using HYPER} \label{appendix:hyper}

\hyper\ is an IDL based software package to perform source extraction and photometry in multi-wavelength surveys \citep{hyper}.  One of the primary advantages of Hyper is that it does photometry in all wavelengths using the same aperture ellipse that is defined at a reference wavelength. This ensures that the same volume of space is sampled at all wavelengths. The size of the ellipse used to do photometry is related to the size of the source (determined by fitting an elliptical Gaussian to the source) by a user supplied aperture factor. However, the choice of aperture factor is challenging when dealing with data where the resolution changes by a significant factor across the different wavelengths. For example, in Hi-GAL data, the resolution varies from 10$''$ to 35$''$ from 70~$\mu$m to 500~$\mu$m. If the aperture factor is too large, then the quality of photometry at short wavelengths will suffer due to contribution from nearby sources (in crowded fields) and diffuse emission. In contrast, if the aperture factor is too small, then a significant fraction of flux can be missed at long wavelengths.

To resolve this problem, point sources of different sizes were simulated (Chatterjee,~R., 2019, Masters thesis) and \hyper\ was run on the simulated images. The simulations took into account the varying resolution at the different wavelengths. Based on simulations of two sources that are closely spaced in a binary, it was determined that the default aperture factor of 2.35 provided the best results in extracting flux densities of both sources. Higher values of the aperture factor resulted in the flux density of the weaker source to be significantly underestimated. Using this choice of aperture factor and a reference wavelength of 250~$\mu$m, the simulations determined the fraction of flux density recovered. This was then used to determine correction factors that need to be applied to the flux densities determined by \hyper\ in order to determine the true flux densities. The correction factors range from 1.04 at 160~$\mu$m to 1.87 at 500~$\mu$m.

\newpage
\section{Submillimeter array images of the target sample}\label{smafigures}
\begin{figure*}[h]
	\centering
	\includegraphics[width=0.8\textwidth]{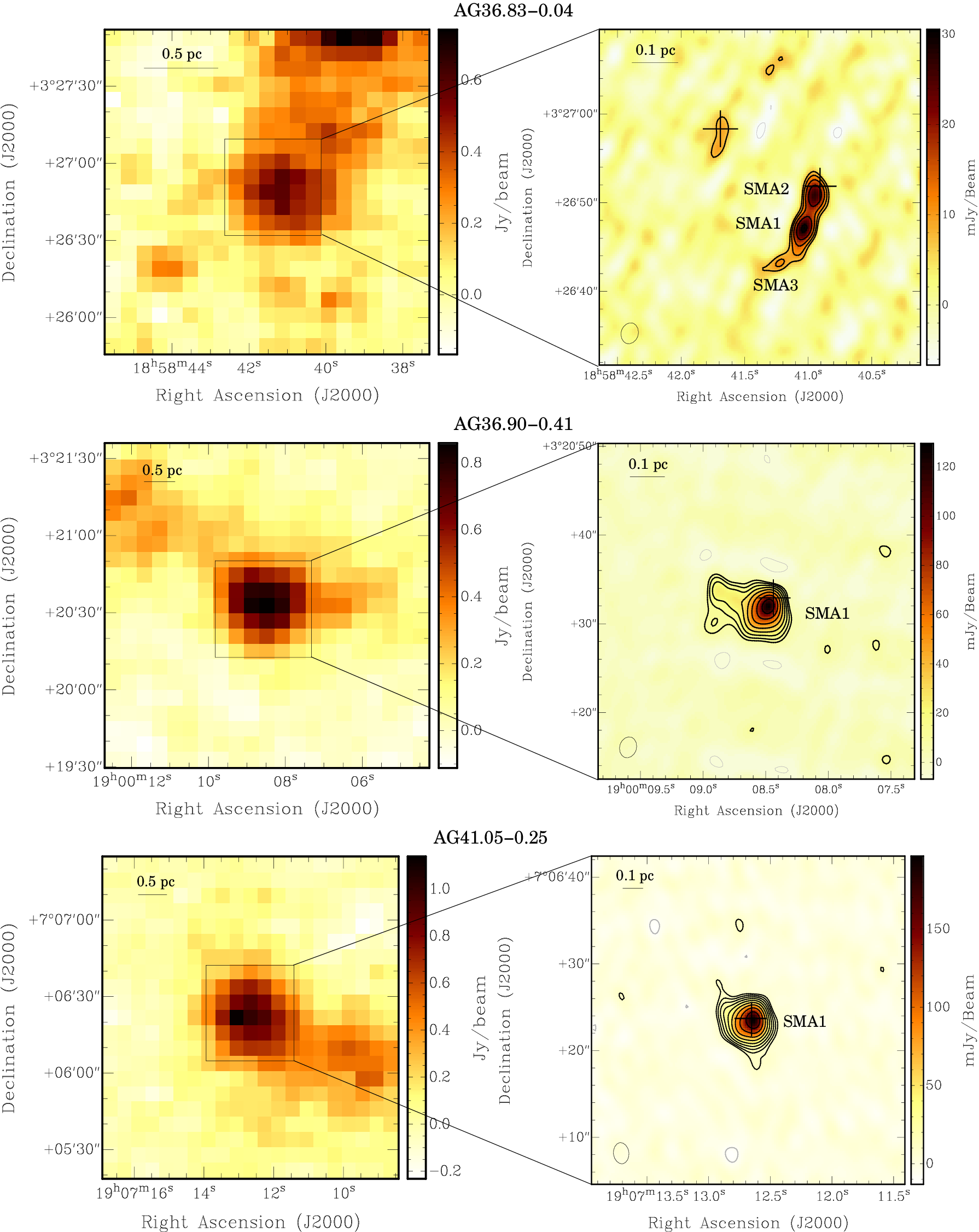}
	\caption{The left panel shows the 870~$\mu$m image from ATLASGAL while the right panel shows the 850~$\mu$m image using the SMA in its compact configuration. The grey contours show emission at the $-3\sigma$ level (see Table~\ref{smaobs}), while the black contours begin at $3\sigma$ and increase in factors of $\sqrt{2}$. The synthesized beam is shown in the bottom left corner of each SMA image. The field of view of the SMA map in the right panel is shown by the box in the left panel. The names of the fragments detected in each source (see Table~\ref{fraglist}) are indicated in the SMA maps. The plus signs indicate the locations of 24~$\mu$m point sources in the MIPSGAL survey.}
\end{figure*}

\begin{figure*}[h]
	\centering
	\figurenum{12}
	\includegraphics[width=0.73\textwidth]{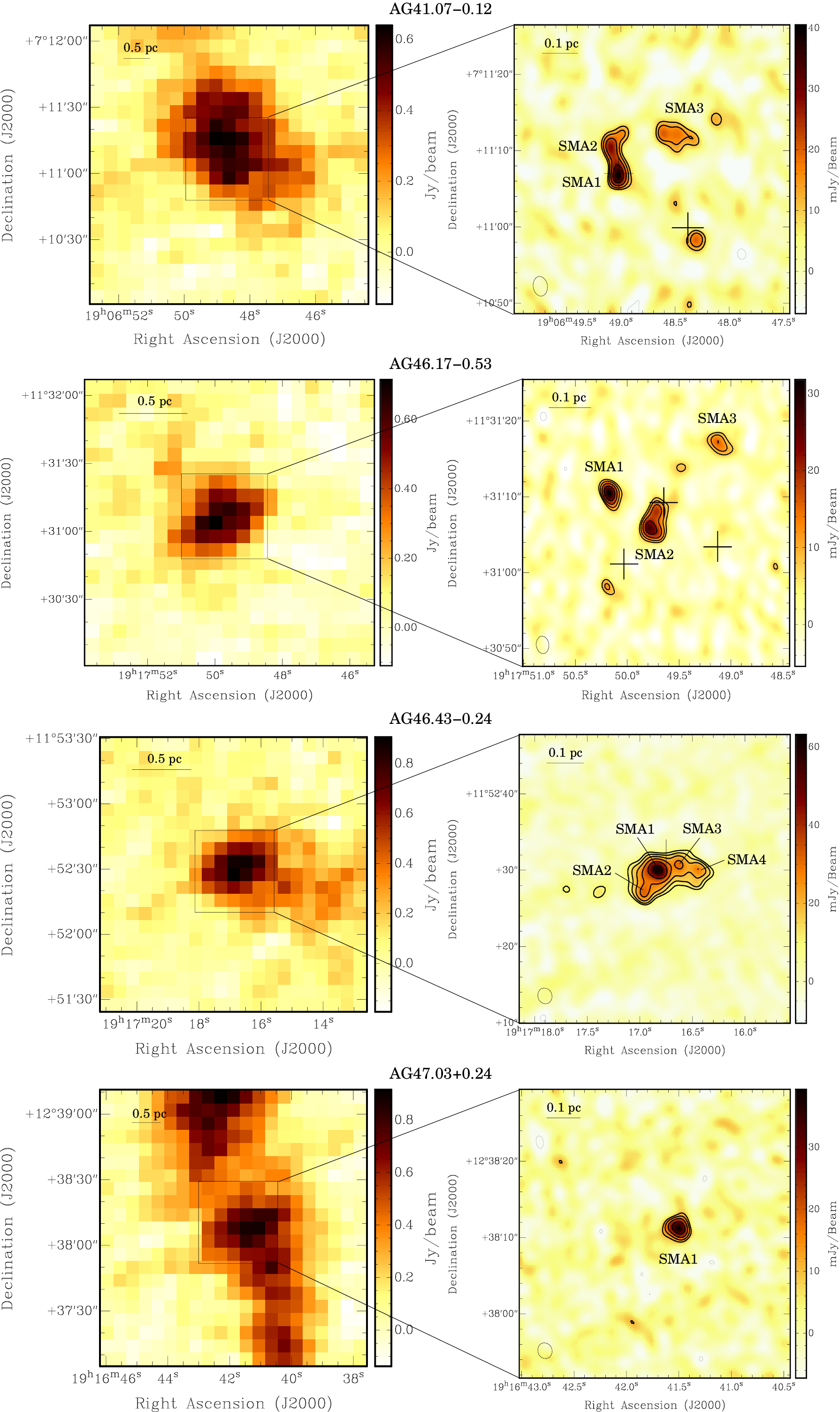}
	\caption{Continued.}
\end{figure*}

\begin{figure*}[h]
	\centering
	\figurenum{12}
	\includegraphics[width=0.73\textwidth]{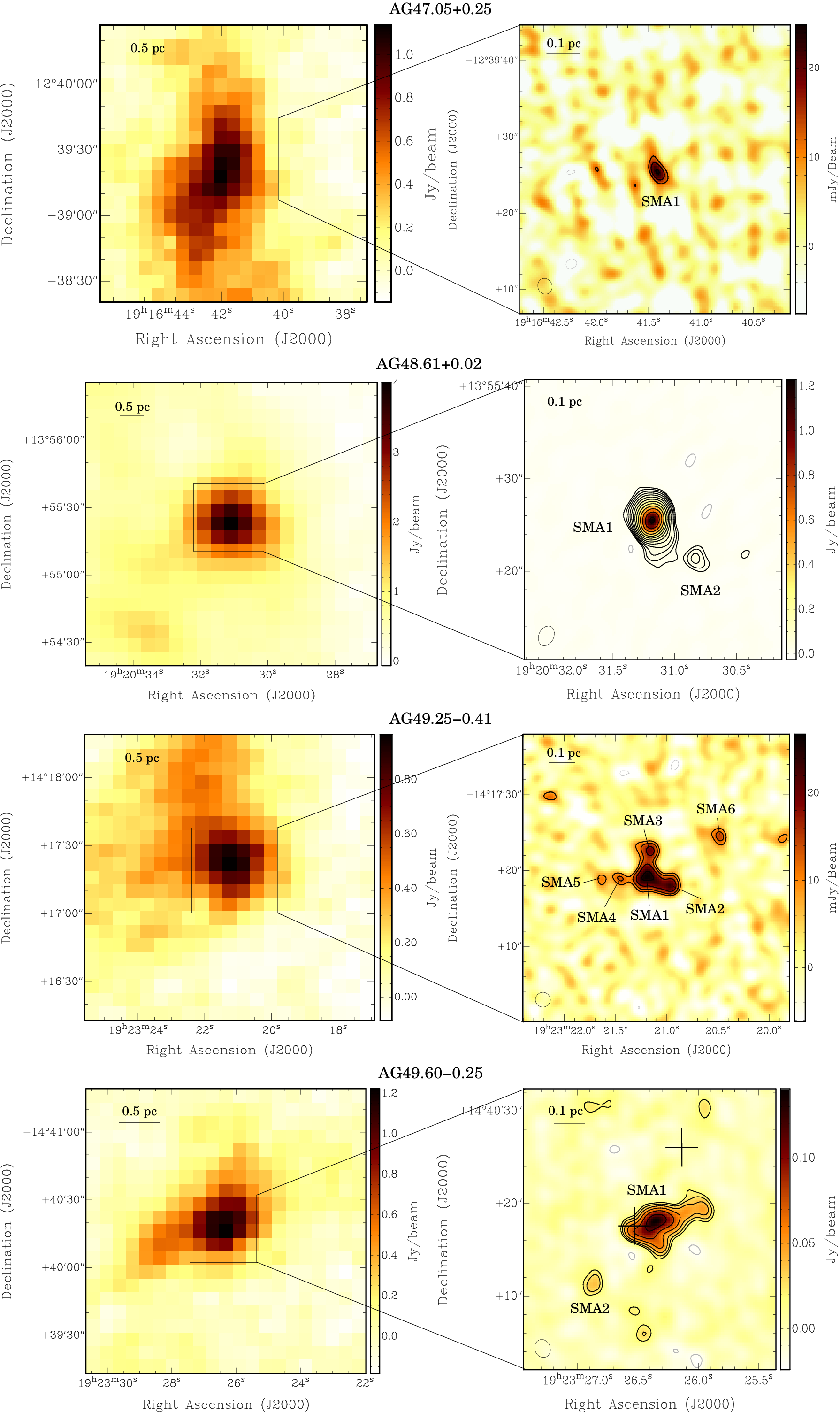}
	\caption{Continued.}
\end{figure*}

\begin{figure*}[h]
	\centering
	\figurenum{12}
	\includegraphics[width=0.73\textwidth]{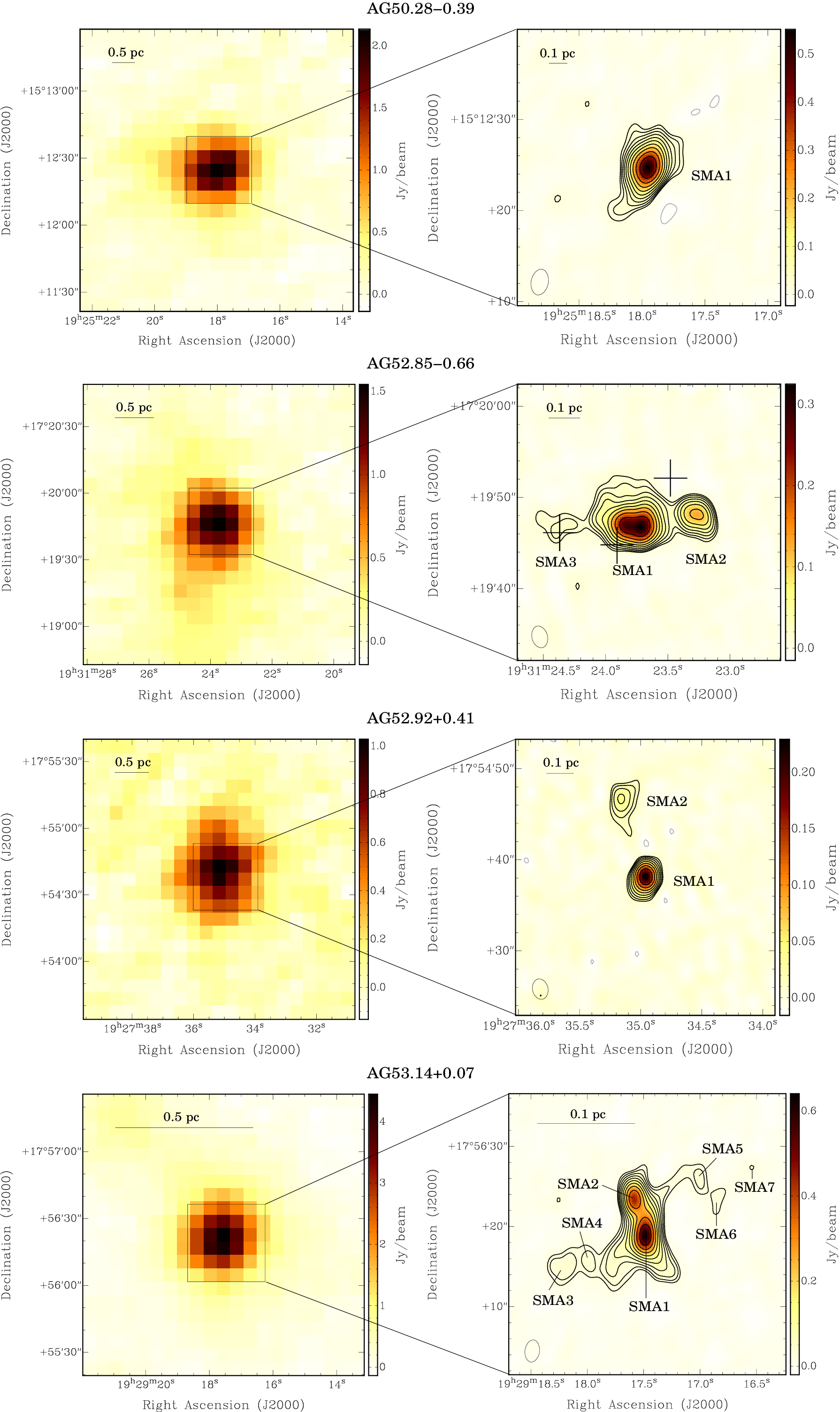}
	\caption{Continued.}
\end{figure*}

\end{document}